\documentclass[twocolumn,times]{aastex61}
\usepackage{epsfig}
\usepackage{amsmath, amsthm, amssymb}
\usepackage{color}
\usepackage{microtype}
\usepackage{float}
\usepackage{verbatim}
\usepackage{wrapfig}
\usepackage[all]{hypcap}

\accepted{January 15, 2018}

\shorttitle{2D CCSNe}
\shortauthors{O'Connor \& Couch}

\def\sref{\S\ref}
\def\fref{Fig.~\ref}

\def\tref{Table~\ref}

\def\eref{Eq.~\ref}
\def\msun{M$_\odot$}
\newcommand\phrv{\rm{Phys. Rev.}}

\newcommand{\code}[1]{\texttt{#1}}

\begin{document}

\title{Two Dimensional Core-Collapse Supernova Explosions Aided by General Relativity with Multidimensional Neutrino Transport}

\author[0000-0002-8228-796X]{Evan P. O'Connor}
\affiliation{Department of Astronomy and Oskar Klein Centre, Stockholm
  University, AlbaNova, SE-106 91 Stockholm, Sweden \href{mailto:evan.oconnor@astro.su.se}{evan.oconnor@astro.su.se}}
\affil{Department of Physics, North Carolina State University, Campus Code 8202, Raleigh, North Carolina, 27695, USA}

\author[0000-0002-5080-5996]{Sean M. Couch}
\affiliation{Department of Physics and Astronomy, Michigan State University, East Lansing, MI 48824, USA; \href{mailto:couch@pa.msu.edu}{couch@pa.msu.edu}}
\affil{Department of Computational Mathematics, Science, and Engineering, Michigan State University, East Lansing, MI 48824}
\affil{National Superconducting Cyclotron Laboratory, Michigan State University, East Lansing, MI 48824, USA}
\affil{Joint Institute for Nuclear Astrophysics-Center for the Evolution of the Elements, Michigan State University, East Lansing, MI 48824, USA}

\begin{abstract}
  We present results from simulations of core-collapse
  supernovae in {\tt FLASH} using a newly-implemented multidimensional
  neutrino transport scheme and a newly-implemented general
  relativistic (GR) treatment of gravity.
  We use a two-moment method with an analytic closure (so-called M1
  transport) for the neutrino transport.  This transport is multienergy, multispecies,
  velocity-dependent and truly multidimensional, i.e., we do not assume
  the commonly used ``ray-by-ray'' approximation.  Our GR gravity is
  implemented in our Newtonian hydrodynamics simulations via an
  effective relativistic potential that closely reproduces the GR
  structure of neutron stars and has been shown to match GR
  simulations of core collapse quite well. In axisymmetry, we simulate
  core-collapse supernovae with four different progenitor models in
  both Newtonian and GR gravity.  We find that the more compact
  protoneutron star structure realized in simulations with GR gravity
  gives higher neutrino luminosities and higher neutrino energies.
  These differences in turn give higher neutrino heating rates
  (upwards of $\sim$20-30\% over the corresponding Newtonian gravity simulations)
  that increase the efficacy of the neutrino mechanism. Three of the
  four models successfully explode in the simulations assuming GREP
  gravity. In our Newtonian gravity simulations, two of the four
  models explode, but at times much later than observed in our GR gravity
  simulations. Our results, both in Newtonian and GR gravity, compare
  well with several other studies in the literature.  These results
  conclusively show that the approximation of Newtonian gravity for
  simulating the core-collapse supernova central engine is not
  acceptable. We also simulate four additional models in GR gravity
  to highlight the growing disparity between parameterized 1D models of
  core-collapse supernovae and the current generation of 2D models.
\end{abstract}


\section{Introduction}
\label{section:intro}

Core-collapse supernovae mark the end stage of stellar evolution for
massive stars.  These explosions are initiated in the cores of stars
with zero-age main sequence masses above 8-10\,$M_\odot$, where
hydrostatic burning of progressively heavier and heavier elements
culminates in the formation of a degenerate and inert iron core with a
mass of at least $\sim$1.2\,$M_\odot$.  The stability of this iron
core against gravitational collapse is due to electron degeneracy
pressure.  Once the iron core grows beyond the effective
Chandrasekhar mass, which can vary depending on the central entropy
and electron fraction \citep{baron:90}, the core begins to collapse.
The collapse is halted when the matter exceeds nuclear density at
which point the repulsive core of the strong nuclear force can supply
enough pressure to stabilize the material once again. The inertia of
the collapsing inner core causes it to overshoot the new hydrostatic
equilibrium, the rebounding of the core is responsible for launching a
shock wave into the still-collapsing outer mantle of the iron
core. For a successful core-collapse supernova, this shock will
ultimately unbind most of the mass of star and help spread the
nucleosynthetic products of stellar evolution throughout the host
galaxy. However, the shock expansion is hampered by two main sinks of
energy. First, the dissociation of the heavy nuclei into neutron and
protons as they pass through the shock gives an inefficient conversion
of the kinetic energy of the infalling material into thermal energy
behind the shock as a modest fraction must go into the internal energy
of the matter.  Second, the temperature behind the shock and the
abundance of free protons now available for electron capture
facilitate energy loss via neutrino emission.  This also removes
thermal energy from behind the shock.  Both of these processes reduce
the thermal pressure behind the shock and bring it into equilibrium
with the ram pressure of the infalling matter.  This causes the shock
to stall and become an accretion shock. The goal of core-collapse
supernova theory over the last $\sim$50 years has been to understand
the shock revival mechanism. The most ubiquitous mechanism has been
the neutrino heating mechanism \citep{colgate:66, bethe:85}.
Neutrino heating is always found in core collapse simulations.  The
intense neutrino radiation field streaming away from the protoneutron
star (PNS) quickly falls out of equilibrium with the matter near the
neutrinosphere. Outside this radius there are lingering
charged-current interactions of electron type neutrinos and
antineutrinos on matter where a net positive transfer of energy
between the neutrinos and the matter occurs. If this energy transfer
is sufficient, it can lead to shock expansion and a successful
explosion.  In spherically symmetric simulations of the collapse of
typical iron cores, the amount of neutrino heating is not sufficient
to drive an explosion \citep{liebendoerfer:01b,rampp:02}. Explosions
in two and three dimensional simulations with energy-dependent neutrino
transport have been reported over the last $\sim$5-10 years
\citep{marek:09, suwa:10, mueller:12a, mueller:12b, takiwaki:12,
  bruenn:13, bruenn:16, nakamura:15, suwa:16, lentz:15, pan:16,
  summa:16, burrows:16}, however in this set of successful explosions
there are discrepancies amongst the results from different codes for
otherwise very similar initial conditions and input physics.

Many of the earliest simulations of core-collapse supernova used or
were concerned with a general relativistic (GR) approach
\citep{colgate:66, may:66,misner:64, wilson:71, bruenn:85,
  burrows:88}, and rightly so since neutron stars are sufficiently
compact that GR dramatically effects their equilibrium
structure. While the importance of including GR gravity in simulations
core-collapse supernovae has always persisted in the literature and is
in use in many current and state-of-the-art multidimensional,
core-collapse supernova calculations \citep{bruenn:13, bruenn:16,
  mueller:12a, marek:09, ott:13a, kuroda:12, lentz:15, skinner:16,
  burrows:16}, many modern simulations have used a purely Newtonian
approximation for gravity \citep{takiwaki:14, nakamura:15, suwa:16,
  couch:14a, handy:14, dolence:15, couch:15a,
  pan:16}. \cite{liebendoerfer:01b} extensively compared GR gravity
and Newtonian gravity in spherical symmetry with a full Boltzmann
neutrino transport solver.  While their baseline simulations in both
prescriptions of gravity fail to explode in 1D, their conclusion is
that, overall, GR is helpful for the development of the core-collapse
supernova explosion.  This conclusion comes out of serendipitous
simulations where incorrect nucleon isoenergetic scattering cross
sections were used.  In these simulations, when GR gravity was used,
the simulation predicted an explosion, but when Newtonian gravity was
used, the simulations failed to achieve explosions.  \cite{bruenn:01, buras:06a}
and \cite{lentz:12a} also compared Newtonian and GR simulations in
spherical symmetry with energy dependent neutrino transport. Like
\cite{liebendoerfer:01b}, they observe higher neutrino luminosities
and energies but do not extensively study the differences, in part
because both simulations still fail to explode owing to the 1D nature
of the simulations.  \cite{kuroda:12} examined the difference between
special relativistic hydrodynamics and GR hydrodynamics in both 1D and
3D core-collapse simulations using approximate neutrino transport (a
grey M1 scheme). However, while the 3D GR simulations show signs of
increased susceptibility to explosion, the simulated time was not long
enough to observe an explosion.  \cite{mueller:12a} also investigated
the influence of Newtonian gravity, GR gravity, and GR effective
potential gravity in 2D core collapse simulations using a variable
Eddington factor, two-moment, energy dependent neutrino transport
scheme and the ray-by-ray+ approximation (where the neutrino transport
is done along radial rays assuming spherical symmetry and with minimal
coupling between neighboring rays). For the classic 15\,$M_\odot$
model from \cite{ww:95}, \cite{mueller:12a} find a late and asymmetric
explosion when using fully GR gravity, but not with Newtonian gravity,
GR effective potential gravity, or with reduced set neutrino
opacities. This was, and still is, the most direct evidence that GR
aids in the explosion mechanism.  It is important to validate the
hypothesis that GR gravity does indeed lead to more favorable
conditions for explosion in modern day simulations, with updated
progenitor models, and loosening the assumption of spherically symmetric.

A number of two-dimensional simulations with similar initial
conditions have been performed by several different groups using
various different treatments of energy-dependent neutrino
transport. The initial conditions, chosen by \cite{bruenn:13},
consists of four models from \cite{woosley:07} with zero-age main
sequence masses of 12\,$M_\odot$, 15\,$M_\odot$, 20\,$M_\odot$, and
25\,$M_\odot$. They employ the EOS of \citet{lseos:91} with
incompressibility parameter $K_0=220$ MeV (hereafter,
LS220). Simulations of these four models using the same EOS were also
presented by \cite{summa:16} and \cite{skinner:16}.
\cite{bruenn:13, summa:16, skinner:16} all use Newtonian
hydrodynamics with a GR effective potential \citep[][Case
A]{marek:06}.  \cite{dolence:15} also simulate core collapse in these
progenitors, however they use the \citet{hshen:11} EOS and assume
purely Newtonian gravity. \cite{suwa:16} and \cite{pan:16} also
simulate at least some of these models, using the LS220 EOS and
Newtonian gravity.  While no collaborative comparison has been done
for these simulations, they do represent the largest set of recent
multidimensional core-collapse supernova models performed by multiple
simulation codes with similar initial conditions. In
\tref{tab:literature}, we summarize the outcome of these simulations
(explosion/no explosion) as well as the time of explosion for
successful models. We divide the results into simulations with a GR
treatment of gravity, and with a Newtonian treatment of gravity.  The
results are quite disparate.  \cite{bruenn:13} find early and
energetic explosions in all four simulated models while
\cite{summa:16} finds explosions in all models, but some explosions
are quite delayed (up to $\sim$500\,ms later than
\citealt{bruenn:13}). Furthermore, \cite{summa:16} conclude that the
final explosion energies will be significantly lower than
\cite{bruenn:13}.  \cite{skinner:16} finds no explosions in all but
one of their models.  For the simulations with Newtonian
gravity, \cite{dolence:15} find no explosions, \cite{suwa:16} finds an
explosion in one model, the 12\,$M_\odot$ progenitor, and
\cite{pan:16} simulate only s15 and s20 but find an explosion in both
models.

In this paper, we perform core collapse supernova simulations for this same set
of progenitors and employ the LS220 EOS
\citep{lseos:91,oconnor:10}. We use a new multidimensional neutrino
transport implementation in the \code{FLASH} simulation
framework \citep{fryxell:00, dubey:09}. We also systematically explore
the differences between core collapse simulations with Newtonian
gravity and with a newly implemented GR effective potential. We
include our results in \tref{tab:literature}.  Our main conclusion is
that a GR treatment of gravity is important and required for
accurately simulating the neutrino mechanism in core-collapse
supernovae. The assumption of Newtonian gravity underestimates the
neutrino luminosity and the mean energies of the emitted spectra,
consequently, Newtonian gravity results in lower neutrino heating
rates in the gain region.  Given the large number of simulation groups
recently using Newtonian gravity to study core-collapse supernovae,
we find it especially important to systematically demonstrate this
point.  We also compare our results to other core-collapse supernova
simulations with comparable initial conditions. Generally, we find our
explosions occur later than those found in the literature. We conclude
these differences are likely due to the differences in the neutrino
opacities and interactions used.

The remainder of this paper is organized as follows.  In
\sref{sec:methods}, we introduce the methods used for computing the GR
effective potential and for our neutrino transport scheme.  Details
and several tests of these methods are presented in the Appendix.  In
\sref{sec:results} we present our main results including a detailed
comparison to publicly available simulation data from
\cite{liebendoerfer:05} and \cite{oconnor:15a}, multidimensional
results assuming Newtonian gravity, and multidimensional results using
GR effective potential gravity. We also briefly discuss the impact of
resolution and random perturbations on our models. We discuss our
Newtonian vs. GREP results in \sref{sec:discussion} as well as compare
our results with those available in the literature for the same
progenitor models and gravity prescriptions. We add four additional
models to our GREP simulations are discuss the results in comparison
to parameterized 1D models of core collapse.  We conclude in
\sref{sec:conclusion}.

\begin{table*}
\caption{}%Literature Results}
\centering
\begin{tabular}{l|c|c|c|c|c|c|c|c|c|c|c}
\hline
Reference & Gravity & EOS & $\nu$ Treatment &\multicolumn{2}{c|}{s12} & \multicolumn{2}{c|}{s15} & \multicolumn{2}{c|}{s20} & \multicolumn{2}{c}{s25} \\
& & & &Exp? & $t_\mathrm{exp}$ [s] &Exp? & $t_\mathrm{exp}$ [s] &Exp? & $t_\mathrm{exp}$ [s] & Exp? & $t_\mathrm{exp}$ [s]\\
\hline
\hline
\cite{bruenn:13} & GREP & LS220 &  MGFLD RxR+ & Yes & 0.236 & Yes & 0.233 & Yes & 0.208 & Yes & 0.212 \\
\cite{summa:16} & GREP & LS220 & VEF RxR+ & Yes & 0.79 & Yes & 0.62 & Yes & 0.32 & Yes & 0.40 \\
this work & GREP & LS220 &  MG M1 & No & -- & Yes & 1.11 & Yes & 0.82& Yes & 0.67\\
\hline
\cite{dolence:15} & NW & H. Shen & MGFLD & No & --& No &  --& No & --& No & --\\
\cite{suwa:16} & NW & LS220 & IDSA RxR & Yes & 0.425 & No & --& No & --& No & --\\
\cite{pan:16} & NW & LS220 & IDSA & N/A & -- & Yes & 0.312& Yes & 0.284& N/A & --\\
this work & NW & LS220 & M1 & No & -- & No & --& Yes  & 1.5 & Yes & 1.1\\
\hline
\end{tabular}
\tablecomments{GREP gravity is used to denote Newtonian hydrodynamic
  simulations with an effective GR potential instead
  of the Newtonian monopole term.  NW gravity is pure Newtonian
  gravity.  The LS220 EOS is the \cite{lseos:91} $K_0=220\,$MeV EOS
  while H. Shen is the EOS from \cite{hshen:11}. The neutrino
  treatment in \cite{bruenn:13} is multigroup flux-limited diffusion (MGFLD), \cite{summa:16} is a two-moment scheme with the closure
  solved by a model Boltzmann equation. These two transport schemes
  use the ray-by-ray+ (RxR+) approximation for the multidimensional
  transport treatment where the transport is solved only in the radial
  direction (along rays). The `+' refers to the addition of advection
  of neutrinos in the lateral direction in optically thick
  regions. \cite{dolence:15} use MGFLD
  as well, but solve the multidimensional transport directly. \cite{suwa:16}
  employ the isotropic diffusion source approximation, akin to MGFLD, and
  use the ray-by-ray approximation. \cite{pan:16} also use IDSA, but
  with a different implementation, including a multidimensional
  diffusion solver. \cite{bruenn:13} defines the explosion time as the
  postbounce time when the shock reaches 500\,km. We use this
  definition for extracting the explosion time from \cite{suwa:16}.  \cite{summa:16} only shows
  shock radius data to 400\,km, therefore we use the postbounce time
  when the shock reaches this radius. However, this makes no qualitative
  difference since the shock expansion is
  quite rapid at this time. We also show our purely Newtonian results.}\label{tab:literature}
\end{table*}

\section{Methods}
\label{sec:methods}

Throughout we use units such that $G=c=1$.

\subsection{Effective General Relativistic Potential}
\label{sec:gravity}

For our implementation of GR gravity, we do not directly solve
Einstein's equations, rather we take an approximate and much simpler
approach. We perform our hydrodynamic simulations using standard
Newtonian hydrodynamics.  However, for the gravitational potential, we
replace the Newtonian monopole contribution to the
potential with an effective GR potential empirically
derived from modified GR structure equations.  The formulation we use
was presented in \cite{keil:97,rampp:02} and later improved by
\cite{marek:06}. In spherical symmetry, the Newtonian
gravitational potential can be determined via the differential
equation,
\begin{equation}
  \frac{d\phi_\mathrm{NW}}{dr} = \frac{m(r)}{r^2}\,,
\end{equation}
where $m(r)$ is the enclosed baryonic mass determined via,
\begin{equation}
\frac{dm}{dr} = 4\pi r^2 \rho\,,\label{eq:NWm}
\end{equation}
where $\rho$ is the mass density.  For Newtonian gravity, the boundary
condition on the enclosed mass is $m(r=0)=0$, while the condition on
the potential is that at the outer edge of our spherical gravitational
potential domain,
\begin{equation}
\phi_\mathrm{NW,bound} = -M_\mathrm{grid}/r_\mathrm{outer}\,.
\end{equation}
\cite{marek:06} modify these equations, using instead,
\begin{eqnarray}
\nonumber
&&\frac{d\phi_\mathrm{eff}}{dr} = \frac{m_\mathrm{TOV}(r) + 4\pi r^3 (P+P_\nu)}{r^2 \Gamma^2}\\
&&\hspace*{2cm}\times \left[\frac{\rho + \rho \epsilon + P}{\rho}\right]\,,\label{eq:phi}
\end{eqnarray}
where $P$, $P_\nu$, and $\epsilon$ are the matter pressure, neutrino
pressure, and specific internal energy, respectively. We take the same
boundary condition as Newtonian gravity,
$\phi_\mathrm{eff,bound} = -M_\mathrm{grid}/r_\mathrm{outer}$.
$\Gamma$ is taken to be,
\begin{equation}
\Gamma = \left [ 1 + v^2 - \frac{2m_\mathrm{TOV}(r)}{r} \right
]^{1/2}\,, \label{eq:gamma}
\end{equation}
where $v^2$ is the square of the average radial velocity and $m_\mathrm{TOV}(r)$ is still the enclosed mass, but rather than
\eref{eq:NWm}, \cite{marek:06} ultimately suggest to take the
enclosed mass as the solution to the following differential equation,
\begin{equation}
\frac{dm_\mathrm{TOV}}{dr} = 4\pi r^2 (\rho + \rho \epsilon +
E_\nu + \frac{v_iF^i_\nu}{\Gamma}) \times\Gamma\,.\label{eq:mtov}
\end{equation}
$m_\mathrm{TOV}$, similar to the Tolman-Oppenheimer-Volkoff (TOV)
mass, accounts for internal matter energy, and in this case, the
energy density and momentum density of the neutrinos ($E_\nu$ and $F^i_\nu$,
respectively). The factor of $\Gamma$, which is not a part of the
standard TOV mass, was empirically found by \cite{marek:06} to better
reproduce actual GR calculations.  This is the `case A' GR effective
potential of \cite{marek:06}. In the Appendix, we perform an unstable
neutron star migration test, in both 1D and 2D, and obtain results
consistent with \cite{marek:06}. We have also tested black hole
formation in our \code{FLASH} implementation using the 40$M_\odot$
progenitor model from \cite{woosley:07}.  We find a black hole
formation time that is $\sim$557\,ms, or $\sim$22\,ms later than the
full GR simulations from \cite{oconnor:15a}. Both of these tests give
us confidence that our implementation of the GR effective potential is
consistent with that of \citet{marek:06} and close to true GR gravity.

For practical implementation in \code{FLASH}, since we use a
cylindrical grid, we must spherically average the matter and neutrino
fields to obtain $\rho$, $P$, $P_\nu$, $\epsilon$, $v$, and $E_\nu$ as
a function of radius. We take the grid center as the center of our
spherical average.  We then numerically solve \eref{eq:mtov} and
\eref{eq:gamma} iteratively as $\Gamma$ depends on $m_\mathrm{TOV}$
and vice versa.  After solving for these terms, we integrate
\eref{eq:phi} to obtain the spherically symmetric gravitational
potential.  For the 2D simulations (both Newtonian gravity and
effective GR gravity) presented here we only use the monopole term for
determining the total gravitational potential.

\subsection{Neutrino Radiation Transport}
We implement a truly multidimensional, two-moment, energy-dependent,
multispecies, neutrino radiation transport scheme with an analytic
closure in {\tt{FLASH}}. Our implementation follows closely that of
\cite{oconnor:15a} which is based on the formalisms of
\cite{cardall:13rad, shibata:11}. In order to incorporate general
relativistic aspects into the transport we follow the assumptions of
\cite{rampp:02}: we retain the $g_{tt}$ term (here it is the lapse,
$\alpha$), but neglect the $g_{rr}$ term.  As in \cite{oconnor:15a},
we distinguish between the laboratory frame where the neutrino moments
are evolved and the fluid frame where the energy of each neutrino bin
is defined and the neutrino interactions take place. We explicitly
couple neighboring energy bins to account for gravitational redshift
and velocity gradient terms that transport neutrinos between energy
bins.  In this work, we ignore inelastic neutrino scattering
processes. With these assumptions, the evolution equations for the zeroth
($E$; neutrino energy density) and first ($F^i$; neutrino momentum
density) energy-dependent neutrino moments in the laboratory frame
are, in spherical symmetry,
\begin{eqnarray}
\nonumber
&&  \partial_t E + \frac{1}{r^2}\partial_r [\alpha r^2 F^r] - \partial_\nu \left[\alpha \nu (L^{rr}\partial_rv_r +
    F^r\partial_r\phi)\right] \\
&&\hspace*{0.2cm}= \alpha (W [\eta -\kappa_a J] -  [\kappa_a + \kappa_s]H^t -  F^r\partial_r \phi)\,,\label{eq:Esph}\\
  \nonumber
 && \partial_t F^r + \frac{1}{r^2}\partial_r [\alpha r^2 P^{rr}]  - \partial_\nu \left[\alpha \nu (N^{rrr}\partial_rv_r +
    P^{rr}\partial_r\phi)\right]\\
\nonumber
&&\hspace*{0.2cm}= \alpha(- [\kappa_s +\kappa_a]H^r +  W [\eta -\kappa_a J]v^r  \\
  &&\hspace*{0.4cm}+ \frac{P^{\theta \theta} + P^{\phi \phi}}{r}-  E \partial_r \phi)\,,
\end{eqnarray}
and cylindrical symmetry,
\begin{eqnarray}
\nonumber
&&\partial_t E + \frac{1}{r}\partial_r [\alpha r F^r] + \partial_z
[\alpha F^z]\\
\nonumber
&&\hspace*{1cm} - \partial_\nu \left[\alpha \nu (L^{ij}\partial_iv_j +
    F^i\partial_i\phi)\right]\\
 &&\hspace*{0.2cm}=\alpha (W [\eta - \kappa_a J] -  [\kappa_a+\kappa_s]H^t -  F^i\partial_i \phi)\,,\\
  \nonumber
&&  \partial_t F^r + \frac{1}{r}\partial_r [\alpha r P^{rr}] + \partial_z [\alpha P^{zr}]\\
\nonumber
&&\hspace*{1cm} - \partial_\nu \left[\alpha \nu (N^{rij}\partial_iv_j +
    P^{ri}\partial_i\phi)\right]\\
\nonumber
&&\hspace*{0.2cm}= -\alpha[\kappa_s +\kappa_a]H^r +  W\alpha [\eta
-\kappa_a J]v^r \\
&&\hspace*{0.4cm}+ \alpha \frac{P^{\phi \phi}}{r} - \alpha E \partial_r \phi \,,\\
  \nonumber
&&  \partial_t F^z + \frac{1}{r}\partial_r [\alpha r P^{rz}]
+ \partial_z [\alpha P^{zz}] \\
\nonumber
&&\hspace*{1cm}- \partial_\nu \left[\alpha \nu (N^{zij}\partial_iv_j +
    P^{zi}\partial_i\phi)\right]\\
&&\hspace*{0.2cm}= \alpha (-[\kappa_s + \kappa_a]H^z +  W [\eta -\kappa_a J]v^z - E \partial_z \phi)\,.
\label{eq:Fzcyl}
\end{eqnarray}
In these equations, $P^{ij}$ is the second moment of the neutrino
distribution function, $L^{ij}$ and $N^{ijk}$ are higher order tensors
used for the explicit energy-space flux calculation.
$W=1/[1-v_iv^i]^{1/2}$ is the Lorentz factor, $\alpha = \exp{(\phi)}$
is the lapse and $\phi$ is the gravitational potential from
\eref{eq:phi}. In our Newtonian simulations these general relativistic
terms do not appear in the moment evolution equations
(i.e. $\alpha=1; \partial_i\phi=0$). $\eta$, $\kappa_a$, and
$\kappa_s$ are the neutrino emissivity, absorption opacity, and
scattering opacity, respectively, which depend on the local density,
temperature, and electron fraction as well as the neutrino species and
energy. $J$ and $H^\alpha$ are the fluid frame energy density and
momentum density and must be written in terms of the laboratory frame
values in order to perform the implicit solve,
\begin{eqnarray}
J &= &W^2[ E - 2 F^i v_i + v^i v^j P_{ij}]\,,\label{eq:fluidframeJ}\\
H^t &=& W^3[ -(E-F^iv_i) v^2 + F^i v_i - v_i v_j P^{ij}]\,,\\
H^i &= &W^3[ -(E -  F^j v_j)v^i + F^i -  v_j P^{ij}]\,.\label{eq:fluidframehi}
\end{eqnarray}

For the specific details of our implemenation, including the methods
used to solve for the closure ($P^{ij}$, $L^{ij}$ and $N^{ijk}$), the
calculation of the explicit fluxes (both in spatial and those in the
energy-space), the calculation of the neutrino interaction
coefficients ($\eta$, $\kappa_a$, and $\kappa_s$), and the methods
used to couple the radiation to the hydrodynamics, we refer the reader
to Appendix~\ref{app:code}.

\subsection{\code{FLASH} simulation setup}
\label{sec:simsetup}

Both the hydrodynamics and the radiation transport share a common time
step that is set by the light crossing time of the smallest grid zone
and a Courant factor. We use a cylindrical grid with adaptive mesh
refinement (AMR) for our simulations. Unless otherwise stated, our
domains extend out to $2\times10^9$\,cm in the radial coordinate, and
$\pm 2\times 10^9$\,cm along the cylindrical axis. Our maximum AMR
block size is $4\times10^8$\,cm and we allow up to 10 total levels of
mesh refinement, giving a minimum AMR block size of $\sim$7.81\,km. We
perform simulations with 16 cells per dimension and per block
resulting in a smallest grid zone with a side width of $\sim$488\,m.
We use a Courant factor of 0.8 that results in a hydrodynamic time
step of $\sim 1.3 \times10^{-6}$\,s. We do two radiation substeps per
hydrodynamic step (see Appendix~\ref{app:code}), therefore the
effective CFL for the radiation substeps is 0.4.  We refine based on a
combination of the second spatial derivatives of the density and
pressure. Unless otherwise stated, at larger radii, we restrict
refinement so as to maintain an effective resolution of at least 0.5
degrees. As a result, we have decrements in refinement at
$\sim$107\,km, $\sim$215\,km, $\sim$430\,km, and so on. Unless
otherwise stated, we use 12 logarithmically-spaced energy groups, with
the highest energy bin $\sim$250\,MeV for our neutrino transport in
2D.  In 1D, we use 18 neutrino energy groups.

We start all \code{FLASH} simulations from \code{GR1D} snapshots at
15\,ms after bounce.  This is to ensure we capture the effect of
inelastic neutrino electron scattering during the collapse phase,
which \code{GR1D} includes but \code{FLASH} does not. We see little
hydrodynamic or radiation effects from this mapping.  Both \code{FLASH}
and \code{GR1D} are using the same nuclear equation of state
implementation, same neutrino transport, same neutrino opacities, and
similar hydrodynamic tools.

\section{Results}
\label{sec:results}

\subsection{s15WW95 - A Detailed Look}
\label{sec:s15WW95}

\begin{figure*}[ht]
\centering
\includegraphics[width=\columnwidth]{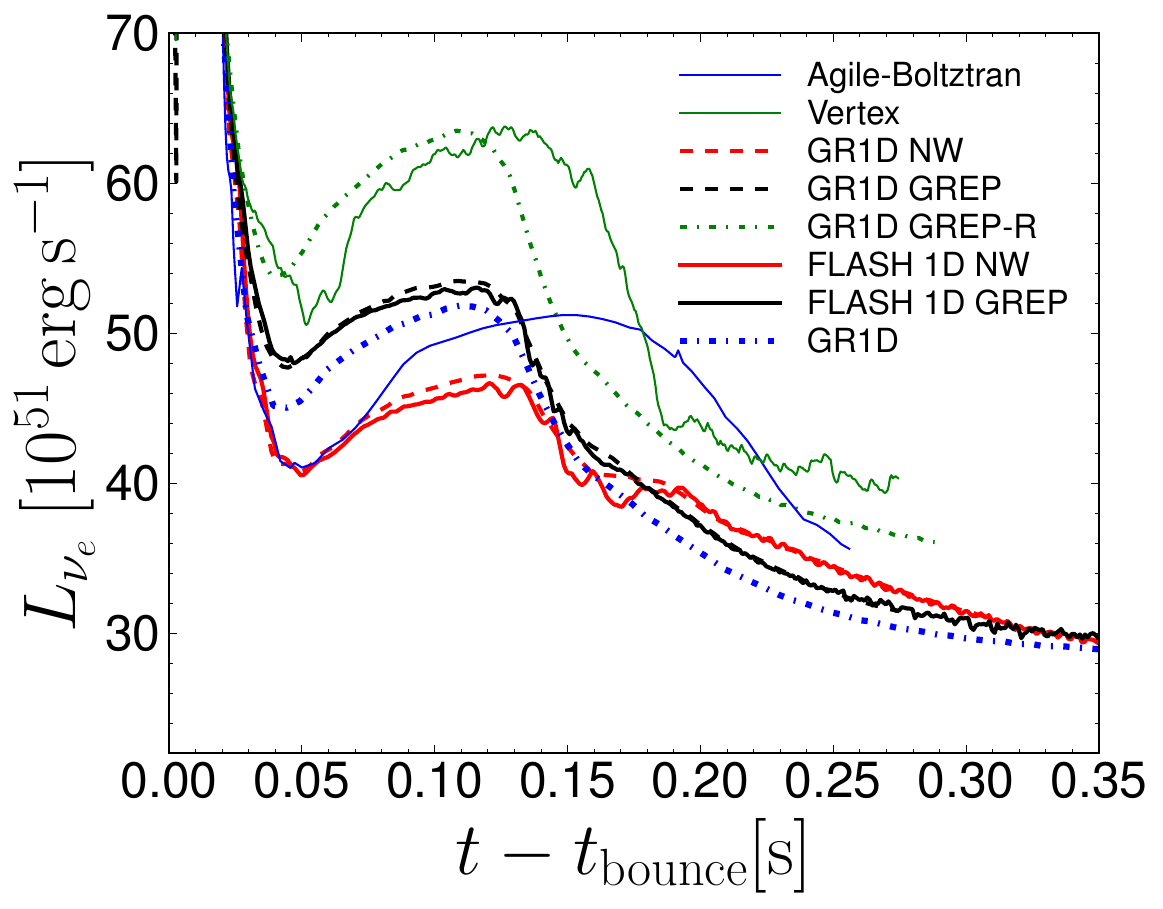}
\includegraphics[width=\columnwidth]{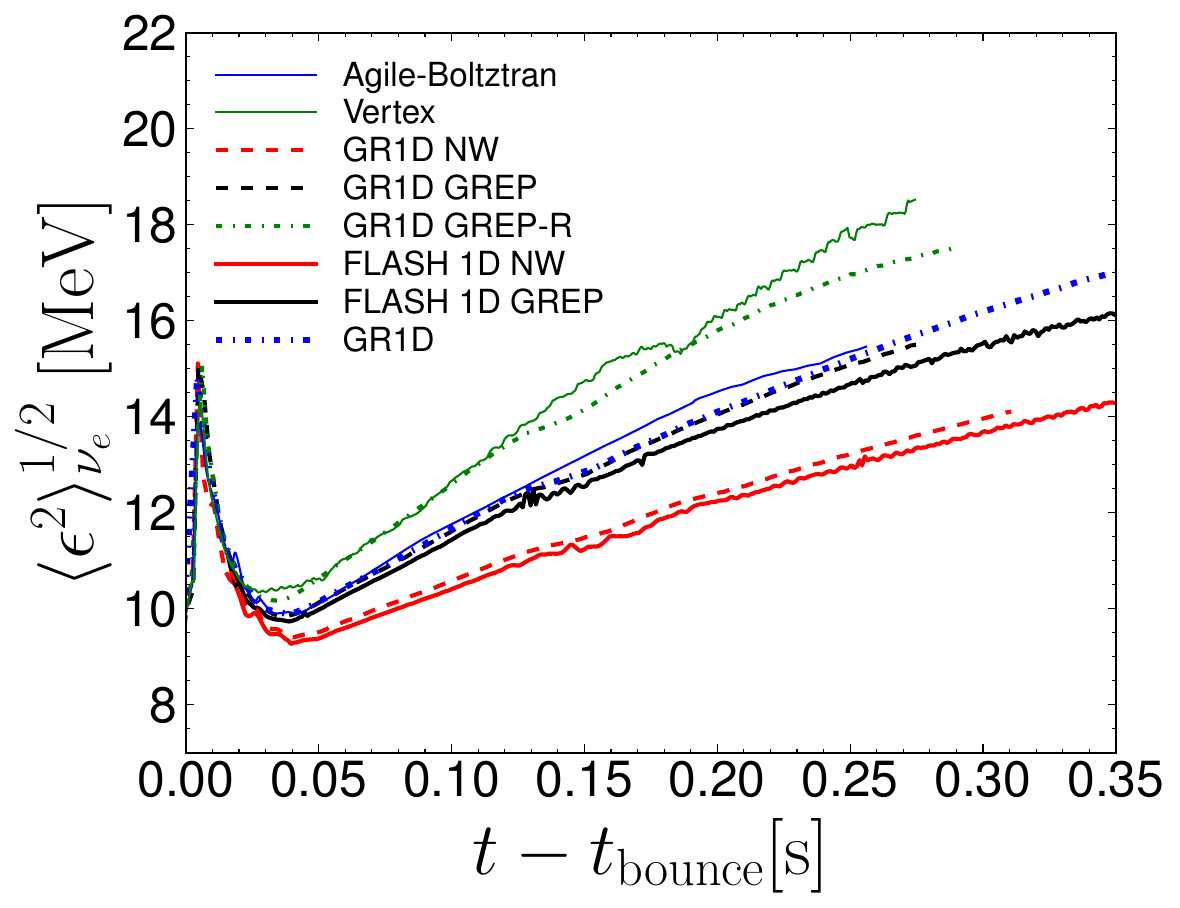}\\
\includegraphics[width=\columnwidth]{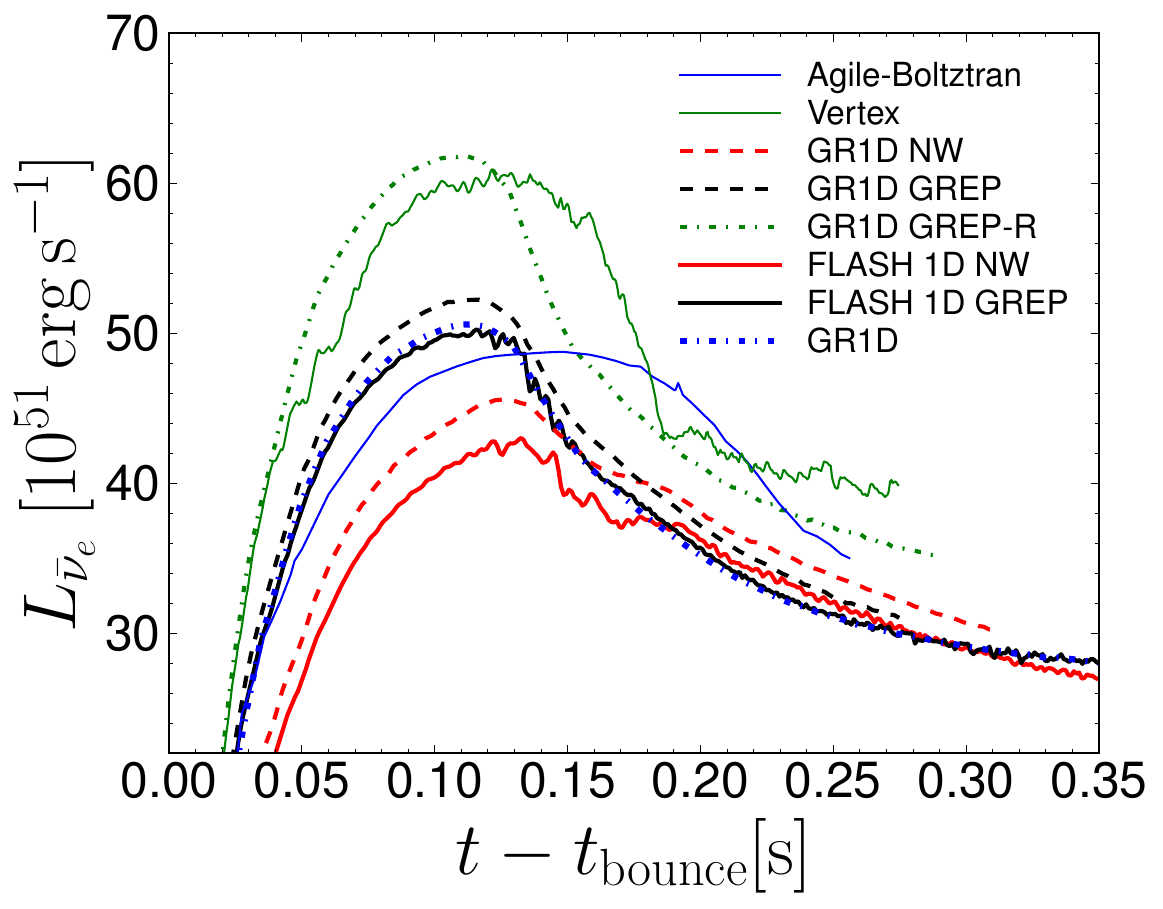}
\includegraphics[width=\columnwidth]{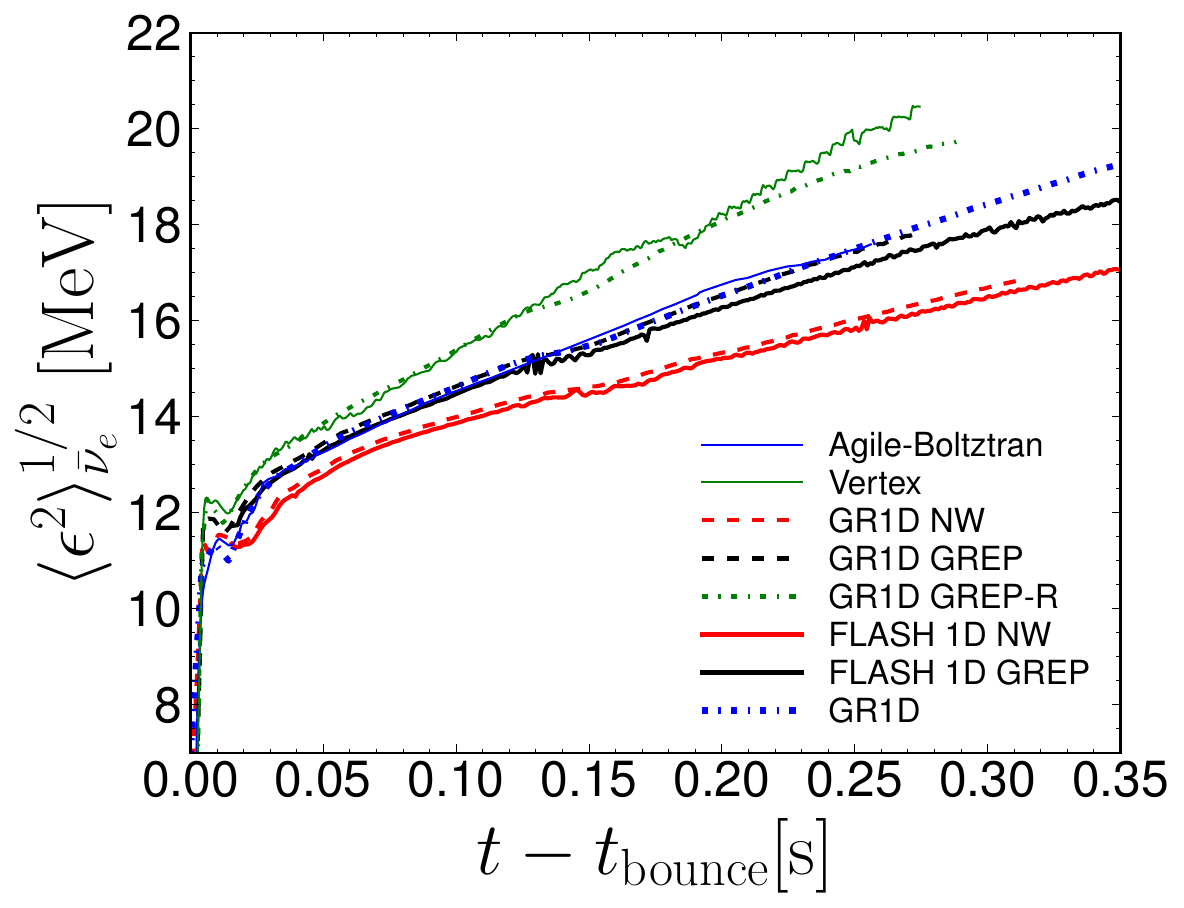}\\
\includegraphics[width=\columnwidth]{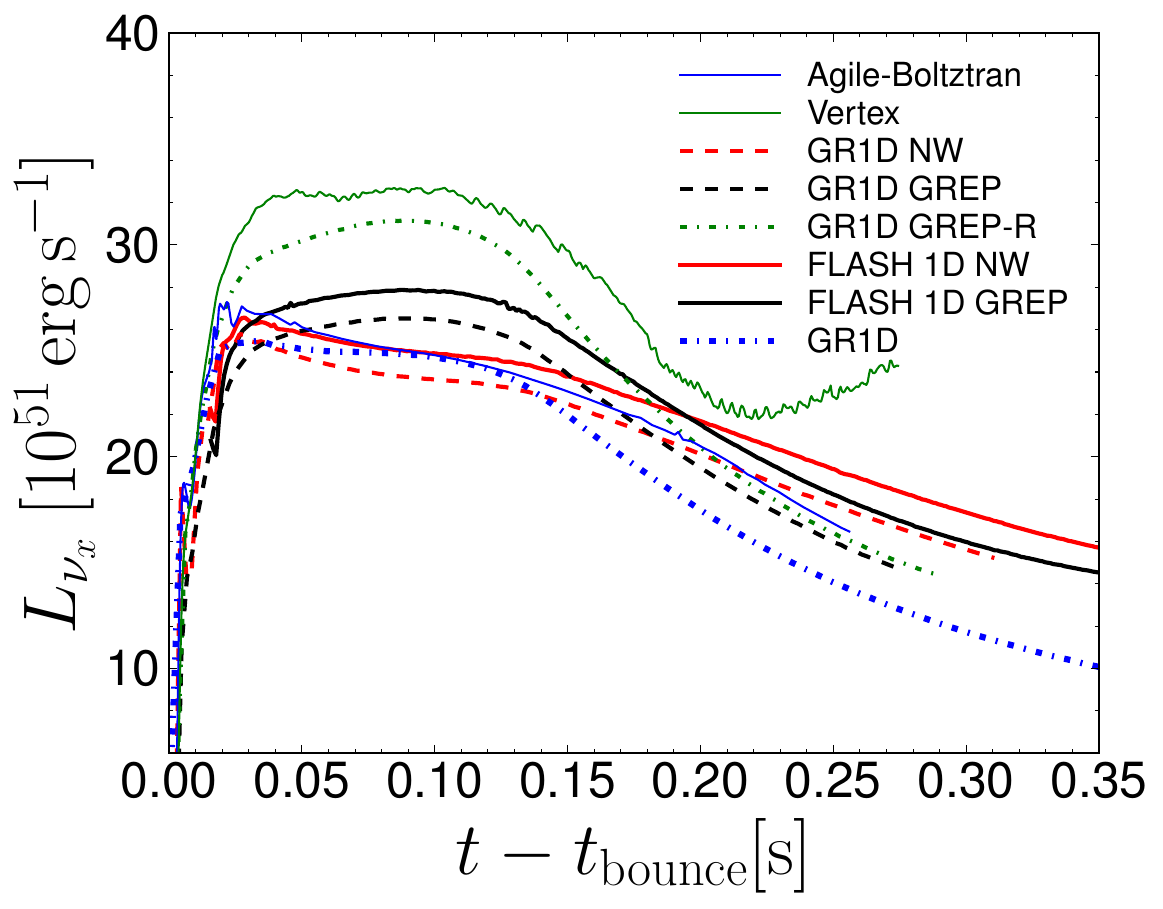}
\includegraphics[width=\columnwidth]{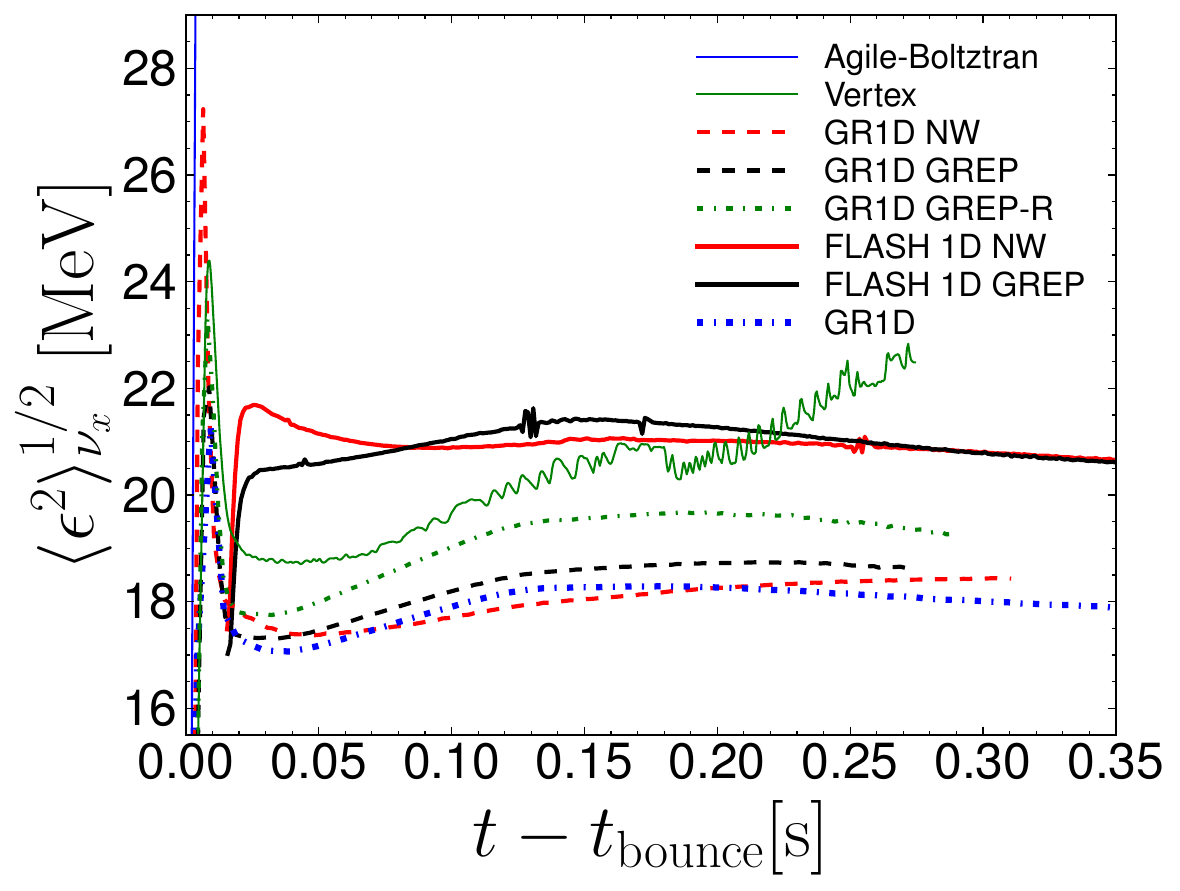}\\
\caption{Neutrino quantities from various calculations of the collapse of
  the s15WW95 progenitor from \cite{ww:95}.  Red lines are Newtonian
  (NW) calculations, black lines are general relativistic calculations
  using the standard `case A' effective potential (denoted GREP) from
  \cite{marek:06}), the green lines are simulations that use the `case
  R' effective potential and are to be compared with the \code{Vertex}
  data from \cite{liebendoerfer:05} (solid green lines), finally the
  dashed-dotted blue lines are from a calculation with \code{GR1D}'s
  GR gravity and should be compared to the
  \emph{Agile}-\code{Boltztran} data from \cite{liebendoerfer:05}
  (solid blue lines).  For the red (NW) and black (GREP) lines, dashed
  lines show data from 1D simulations with \code{GR1D} while the solid
  lines show data from spherically symmetric simulations with
  \code{FLASH}.  All \code{GR1D} simulations include inelastic
  scattering while all \code{FLASH} simulations ignore this neutrino
  process.  The \code{FLASH} simulations are started from the
  corresponding \code{GR1D} run at 15\,ms after core bounce. Note, all
  luminosities and root mean squared energies are given in the fluid
  frame.}\label{fig:s15WW95_comparenu}
\end{figure*}

\begin{figure}[ht]
\centering
\includegraphics[width=0.94\columnwidth]{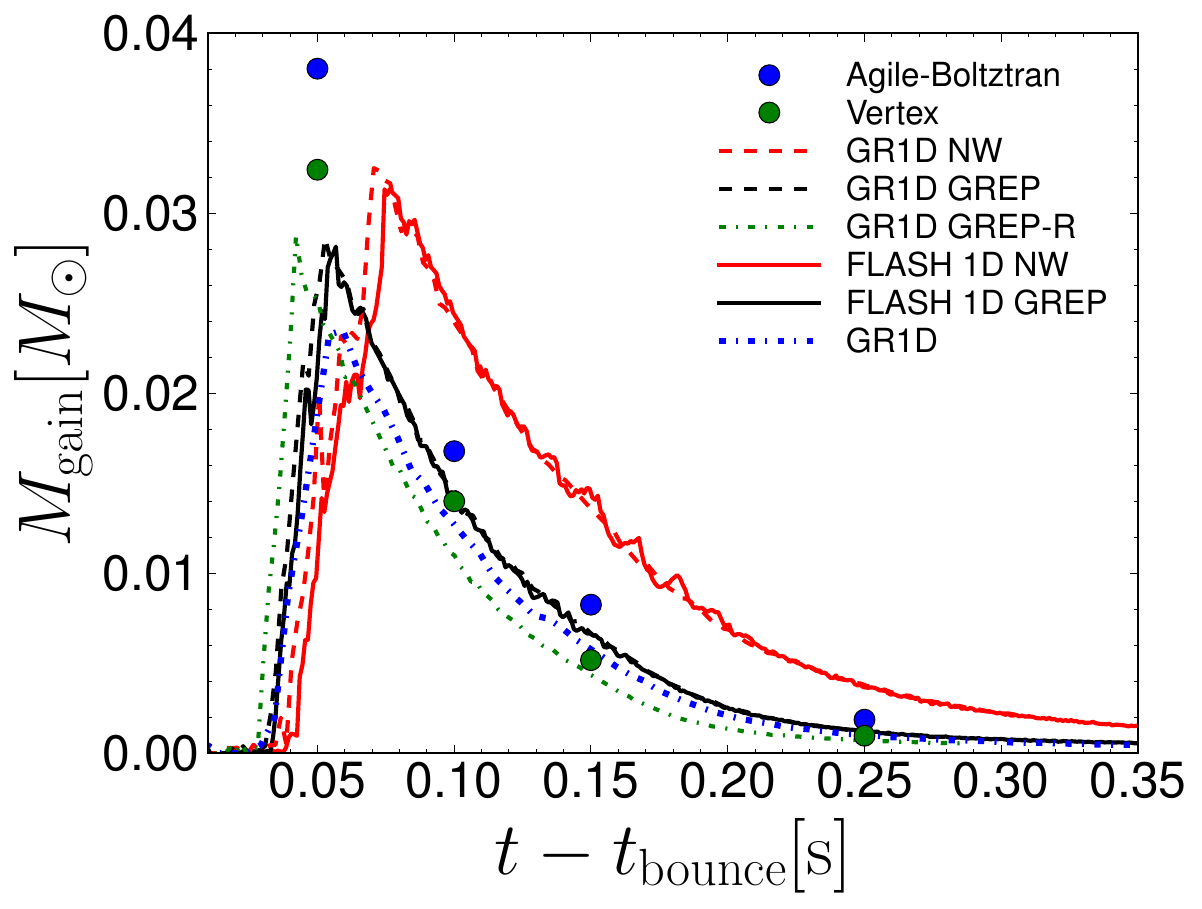}\\
\includegraphics[width=0.94\columnwidth]{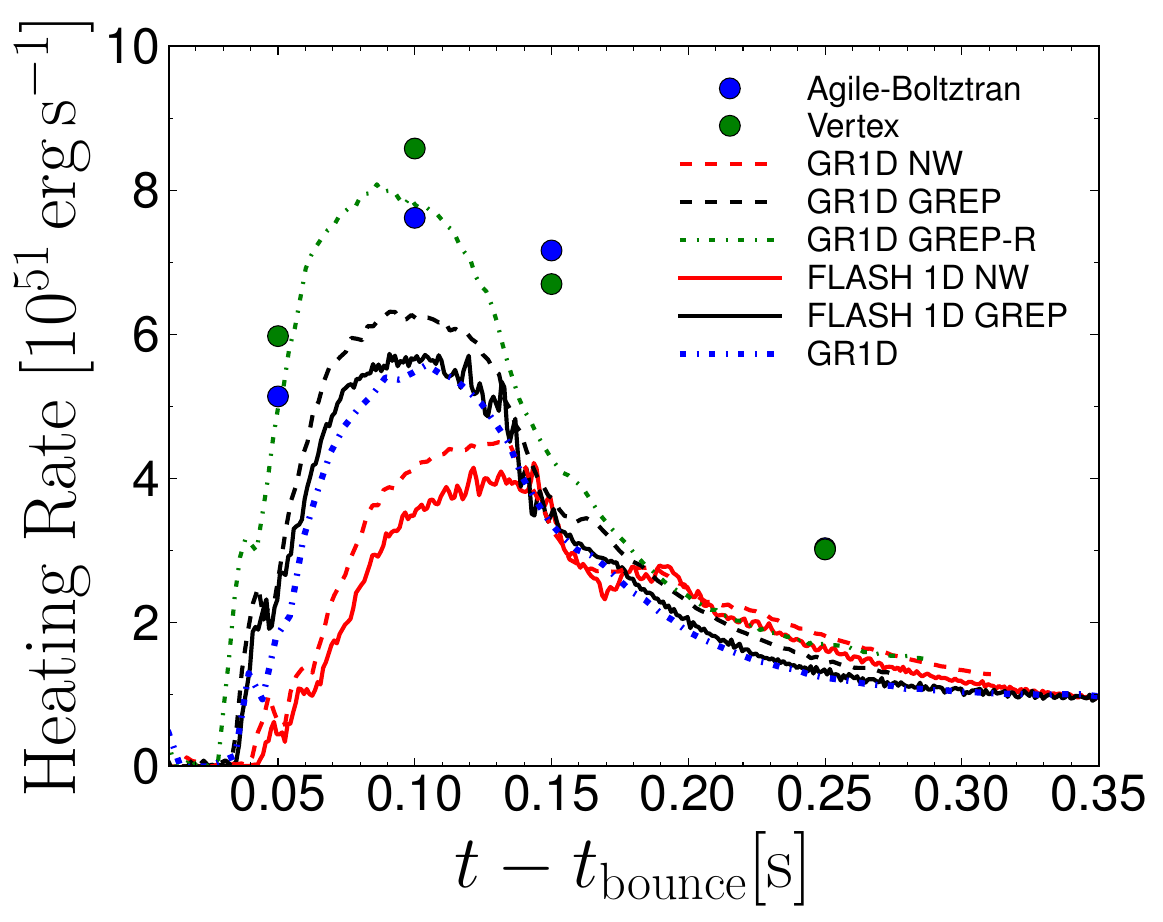}\\
\includegraphics[width=0.94\columnwidth]{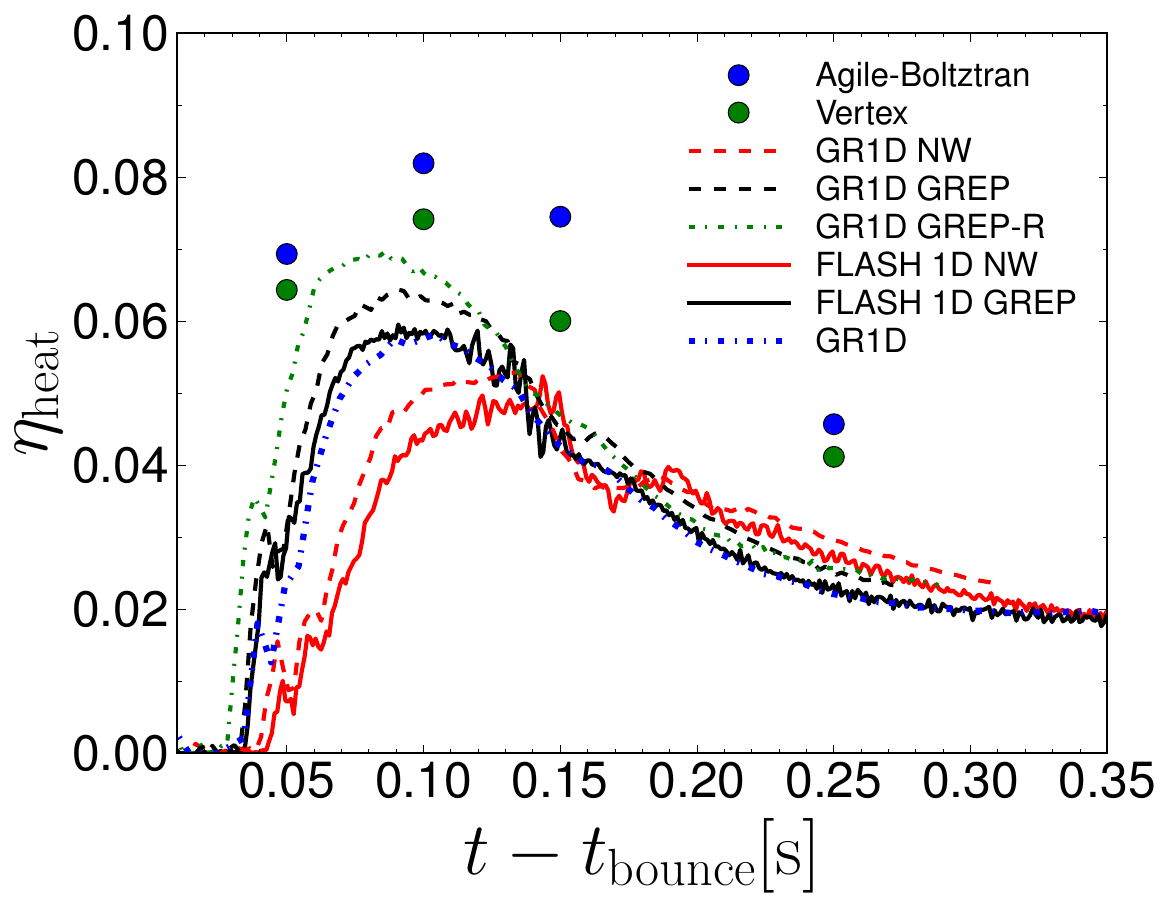}\\
\caption{Gain region data from various calculations of the collapse of
  the s15WW95 progenitor from \cite{ww:95}.  Line and color
  descriptions are the same as \fref{fig:s15WW95_comparenu}. We show
  the mass of the gain region in the top panel, heating rate in the
  gain region in the middle panel, and the heating efficiency defined
  by \eref{eq:etaheat} in the bottom panel. For the
  \emph{Agile}-\code{Boltztran} and \code{Vertex} simulations from
  \cite{liebendoerfer:05}, we can only show data points at select
  times (50, 100, 150, and 250\,ms).}\label{fig:s15WW95_comparegain}
\end{figure}

We use the 15 $M_\odot$ core-collapse supernova progenitor profile
from \cite{ww:95} as a test case, denoted throughout as s15WW95.  To
make comparisons with previous publicly available work
\citep{liebendoerfer:05}, we use the Lattimer \& Swesty EOS with an
incompressibility of 180\,MeV (LS180; \citealt{lseos:91}).  For this
test only, following \cite{liebendoerfer:05}, we do not include weak
magnetism and recoil corrections for the neutrino opacities.  We
present the results in \fref{fig:s15WW95_comparenu} and
\fref{fig:s15WW95_comparegain}. For both figures and in all panels,
red lines are collapse calculations using Newtonian gravity, black
lines are calculations using our standard general relativistic
effective potential, `case A', from \cite{marek:06}. The green lines
(thin dashed-dotted) in uses `case R' from \cite{marek:06} and are
presented for comparison purposes with the data extracted from the
publicly available \code{Vertex} data presented in
\cite{liebendoerfer:05}.  Finally, the blue (thick dashed-dotted) line
is from a fully general relativistic calculation using \code{GR1D}
(data is from \citealt{oconnor:15a}) while the solid blue line and
blue dots are data taken from publicly available
\emph{Agile}-\code{Boltztran} data from \cite{liebendoerfer:05}. For
the red and black lines, the thin dashed lines correspond to data
taken from \code{GR1D} and the thick dashed lines are from
\code{FLASH}. We note that all of the \code{GR1D} simulations (both
GREP and full GR) include inelastic scattering of neutrinos on
electrons, while \code{FLASH} does not. This aspect is especially
important during the collapse phase, hence, we start all \code{FLASH}
simulations from \code{GR1D} profiles at $\sim$15\,ms after
bounce. For our s15WW95 \code{FLASH} simulations the outer extent of
our domain is taken to be $10^9$\,cm.

\fref{fig:s15WW95_comparenu} shows the fluid-frame neutrino
luminosity and root mean squared energy, respectively, of electron
neutrinos (top panels), electron antineutrinos (middle panels), and
heavy-lepton neutrinos (bottom panels) extracted from the simulations
at 500\,km.  \fref{fig:s15WW95_comparegain} shows the
mass in the gain region (top), the total heating rate in the gain
region (middle) and heating efficiency (bottom).  The gain region is
defined as the region within the shock, but outside the
neutrinosphere, where there is an increasing internal energy of the
matter due to neutrino interactions.  The heating efficiency is
defined as the ratio of the rate of energy deposition into the gain
region ($\dot{Q}^\mathrm{gain}_\mathrm{heat}$) to the total
electron-type neutrino luminosity at the gain radius.  Since the
latter is difficult to estimate in multidimensional simulations, we
estimate it by adding the heating rate to the lab-frame neutrino
luminosity extracted from our simulations,
\begin{equation}
\eta_\mathrm{heat} = \frac{\dot{Q}^\mathrm{gain}_\mathrm{heat}}{L^\mathrm{lab}_{\nu_e} +
 L^\mathrm{lab}_{\bar{\nu}_e} + \dot{Q}^\mathrm{gain}_\mathrm{heat}}\,.\label{eq:etaheat}
\end{equation}
For the data from \cite{liebendoerfer:05}, we project their fluid
frame luminosity to the lab frame via $L^\mathrm{lab}_\nu \sim
L^\mathrm{fluid}_\nu (1+v/c)/(1-v/c)$.

Generally, the deeper gravitational well of the general relativistic
simulations (represented by blue, black, and green lines)
allows more gravitational binding energy to be released during the
accretion phase and results in a higher electron neutrino and electron
antineutrino luminosity, which are predominately fueled by accretion,
when compared to the Newtonian simulations (red lines). The deeper
gravitational well also leads to higher matter temperatures at the
neutrinospheres and therefore higher root mean squared energies.  This
gives larger heating rates and heating efficiencies.  The `case R'
effective gravitational potential (labeled as \code{GR1D} GREP-R;
green curve) overestimates the correction to the gravitational
potential and gives even higher luminosities and energies
\citep{marek:06}.  Important for our discussion is that our `case R'
results (green lines) closely match the `case R' results from the
\code{Vertex} code \citep{liebendoerfer:05} This agreement, along with
the effective potential test cases in Appendix \ref{app:GR}, gives us
confidence in our effective potential implementation.  The improved
effective potential presented by \cite{marek:06}, (their `case A'),
comes much closer to reproducing the neutrino luminosities, and
matches very closely the root mean squared energies, of the full GR
codes \emph{Agile}-\code{Boltztran} and \code{GR1D}.

The differences that are seen between the \code{GR1D} and
\code{GR1D}-GREP collapse results are due to the differences between
the Newtonian hydrodynamics + GR effective potential gravity and GR
hydrodynamics + GR gravity.  These differences are also on the order
of the differences between the \code{FLASH} simulations and the
\code{GR1D}-GREP simulations, which differ on the hydrodynamics. With
the exception of the heavy-lepton neutrino quantities, which are
discussed below, the various calculations differ by $\sim2-5\%$ in the
neutrino luminosities and root mean square energies.  The differences
in the heating related quantities are somewhat larger, $\sim 10\%$,
owing to their non-linearly dependence on the neutrino field.

The largest difference between the \code{GR1D} and \code{FLASH}
results is seen in the heavy-lepton sector which is a direct result of
the neglect of inelastic neutrino-electron scattering (NES).  After
bounce, the main effect of NES is to down-scatter neutrinos
\citep{thompson:03}.  Electron-type neutrino and antineutrinos have
charged-current interactions that keep those neutrinos
thermodynamically coupled to the matter until low densities where NES
is not as effective.  However, heavy-lepton neutrinos, which lack
these charged-current interactions, experience a large scattering
dominated region where NES plays an important role. The \code{FLASH}
simulations, both GREP and NW, have on average 4-6\,MeV higher
($\sim25\%$) heavy-lepton $\nu_x$ root mean squared energy and
$\sim2\times10^{51}$\,erg\,s$^{-1}$ ($\sim10\%$) higher
luminosities, consistent with the observations of
\cite{thompson:03}. However, within the time simulated, the different
$\nu_x$ luminosities or the additional heating at higher densities supplied
by inelastic NES does not have an influence on the electron-type
neutrino luminosities or energies, and therefore the heating
quantities.  For the purposes of neutrino signal prediction, this
assumption is quite severe, however there is little effect on the
early core-collapse supernova dynamics.

\subsection{2D Newtonian Results}
\label{sec:NWresults}

\begin{figure*}[ht]
\centering
\includegraphics[width=0.99\columnwidth]{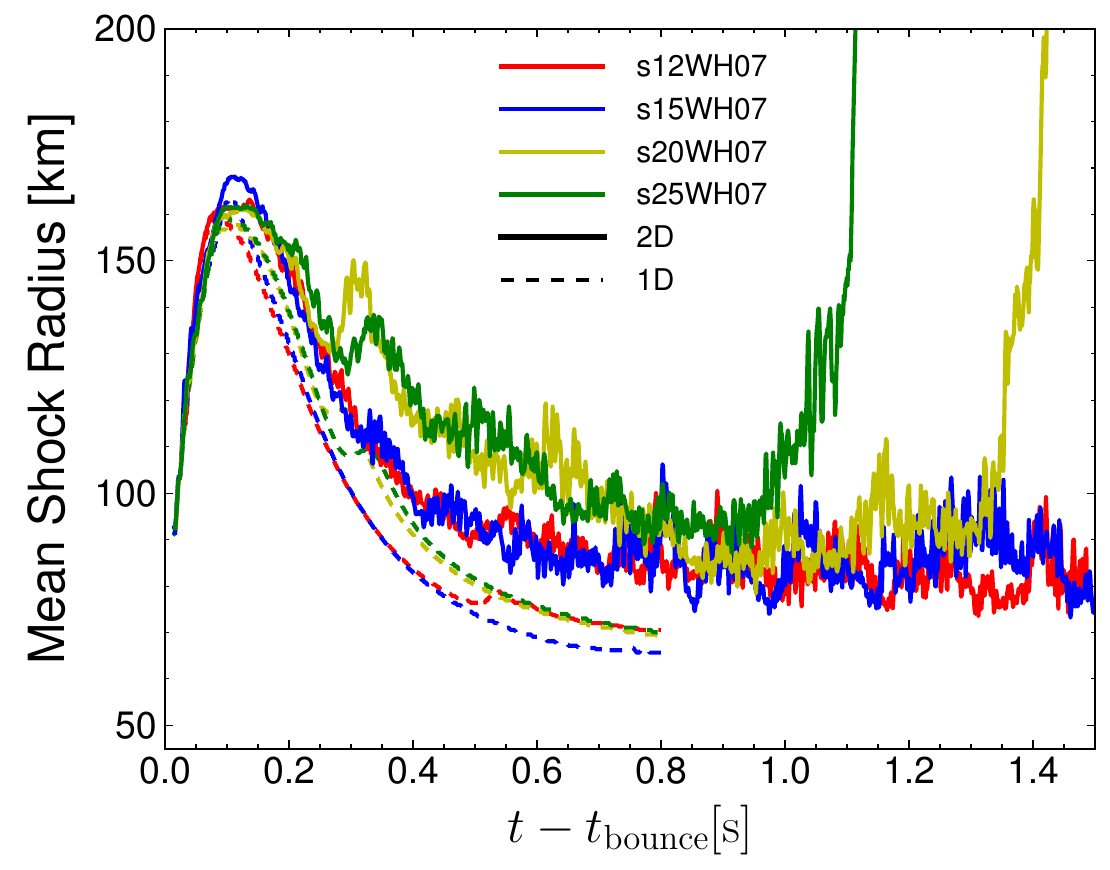}
\includegraphics[width=0.99\columnwidth]{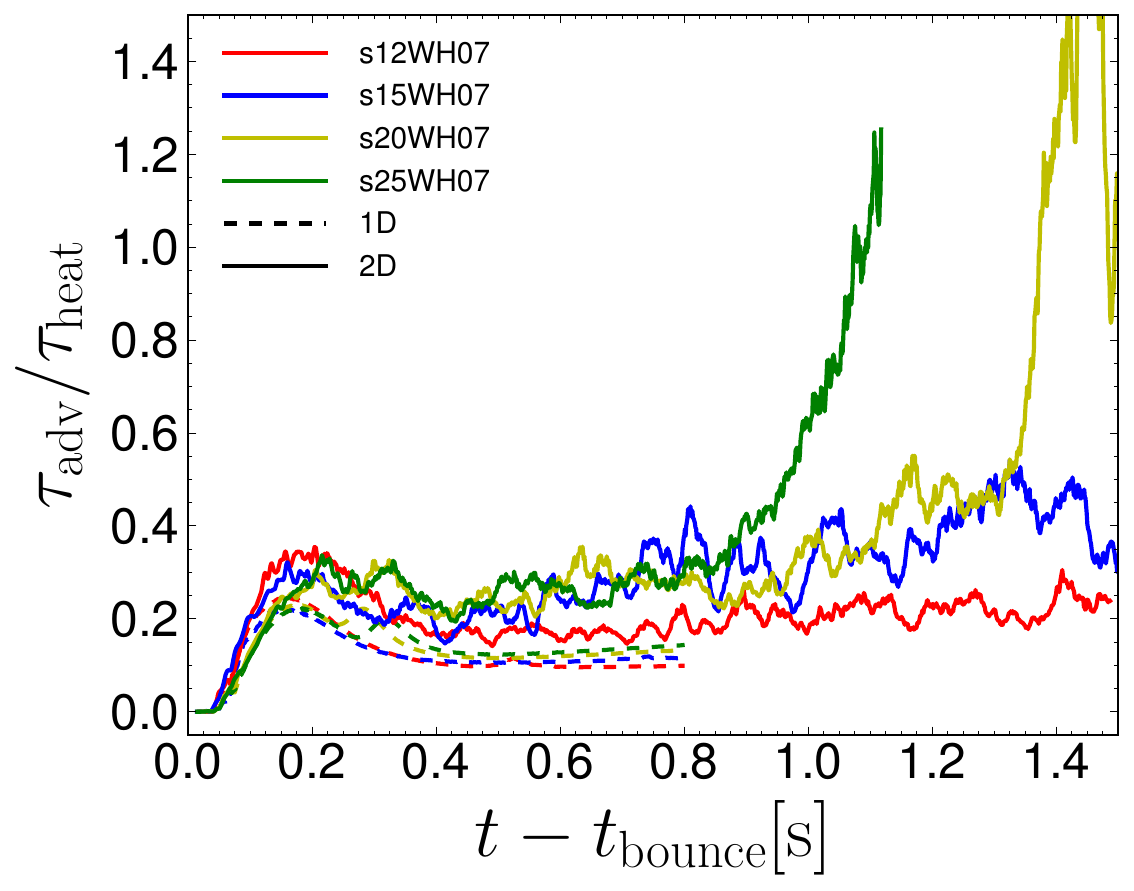}\\
\caption{Simulation results from 1D and 2D core collapse using
  Newtonian gravity in four models. We show the time evolution of the
  mean shock radius (left); $\tau_\mathrm{adv}/\tau_\mathrm{heat}$
  (right).
  Solid lines are from 2D simulations, while dashed lines are from 1D
  simulations. To reduce the scatter, we average
  $\tau_\mathrm{adv}/\tau_\mathrm{heat}$ over a 5\,ms
  window. }\label{fig:2DNewtonian}
\end{figure*}

We perform a total of six two-dimensional core collapse simulations
using Newtonian gravity.  For all simulations, we start from profiles
taken from \code{GR1D} 15\,ms after bounce.  We perform four
simulations of the models introduced in \tref{tab:literature}, s12,
s15, s20, and s25 from \cite{woosley:07} (WH07).  We use our base
resolution and the LS220 EOS.  As mentioned in \sref{sec:gravity}, we
assume $l_\mathrm{max}=0$ in our multipole gravity solver.  We have
tested this assumption by performing a Newtonian transport simulation
with model s12 and with $l_\mathrm{max}=24$. Finally, we test our
effective potential solver by performing a Newtonian simulation with
the s12 progenitor and where the gravity is computed via the algorithm
in \sref{sec:gravity}, but replacing the differential equations with
the Newtonian equivalents.  Theoretically, this is precisely the same
as the standard monopole term for the gravity, but numerically
calculated in a different way. For both of these latter two test
simulations we see no differences other than the stochastic behavior
of the convection and turbulence.

In \fref{fig:2DNewtonian}, we show a subset of results from our
Newtonian simulations. We see a very late time ($\gtrsim 1\,$s)
explosion in two models, s20 and s25, and failure in the other models,
s12 and s15.  Other than the late time explosion, the simulations of
the four models proceed similarly.  In both 1D and 2D, in all four
models, the shock initially expands out and reaches $\sim$160-170\,km
at $\sim$100\,ms after bounce.  Only at this time do the 1D and 2D
simulations begin to depart from each other as convection begins to
strongly develop behind the shock and provide additional pressure
support for the supernova shock \citep{murphy:13, couch:15a}. In 1D
and 2D the shocks begin to recede.  Compositional interfaces reach the
shock front at $\sim$250\,ms, $\sim$300\,ms, and $\sim$500\,ms for
models s20, s25, and s12, respectively. At these times, especially in
models s25 and s20, we see a short phase of shock expansion. The shock
expansion is more prominent in 2D than in 1D. However, this is not
sufficient to reenergize the shock in these simulations. At late times
the accretion rate has dropped enough to drive an explosion. However,
we caution that at these times the baryonic mass of the PNSs is
approaching 2.2\,$M_\odot$.  The Newtonian gravity approximation is
expected to be far from valid in this regime as the maximum baryonic
mass of the LS220 EOS $2.41\,M_\odot$.

A useful diagnostic of proximity to explosion is the ratio of the
advection time through the gain region to the time it takes to unbind
the gain material through neutrino heating.  A ratio of 1 can be
thought of as a runaway condition where matter is unbound before it
can be advected into the cooling region.  Quantitatively we define
this ratio via,
\begin{equation}
\frac{\tau_\mathrm{adv}}{\tau_\mathrm{heat}} =
\frac{M_\mathrm{gain}/\dot{M}}{|E_\mathrm{gain}|/\dot{Q}_\mathrm{heat}}\,.\label{eq:tadvovertheat}
\end{equation}
We define the gain region to consist of any grid zone with a net
positive change in the internal energy of the matter due to neutrino
interactions.  $M_\mathrm{gain}$ is the total mass of those grid
zones, $\dot{M}$ is the accretion rate measured at 500\,km,
$\dot{Q}_\mathrm{heat}$ is the rate of energy exchange between the
neutrino and the matter in the gain region, and
$E_\mathrm{gain} = \sum_\mathrm{gain} m_i [1/2|v_i|^2 + \epsilon_i +
\phi_i]$\,,
is the energy of the matter in the gain region.  For the latter, $v_i$
is the matter velocity, $\phi_i$ is the gravitational potential, and
$\epsilon_i$ is the specific internal energy of the matter.  For
consistency with other works, we take the zero point defined in LS220
EOS for $\epsilon_i$. In the top right panel of \fref{fig:2DNewtonian}
we show this quantity for our Newtonian models. For most of the time,
the WH07 models we simulate are well below
$\tau_\mathrm{adv}/\tau_\mathrm{heat} \sim 1$. Only at late times does
this ratio approach 1 for the two exploding models.

\begin{figure}[htbp]
\centering
\includegraphics[width=0.94\columnwidth]{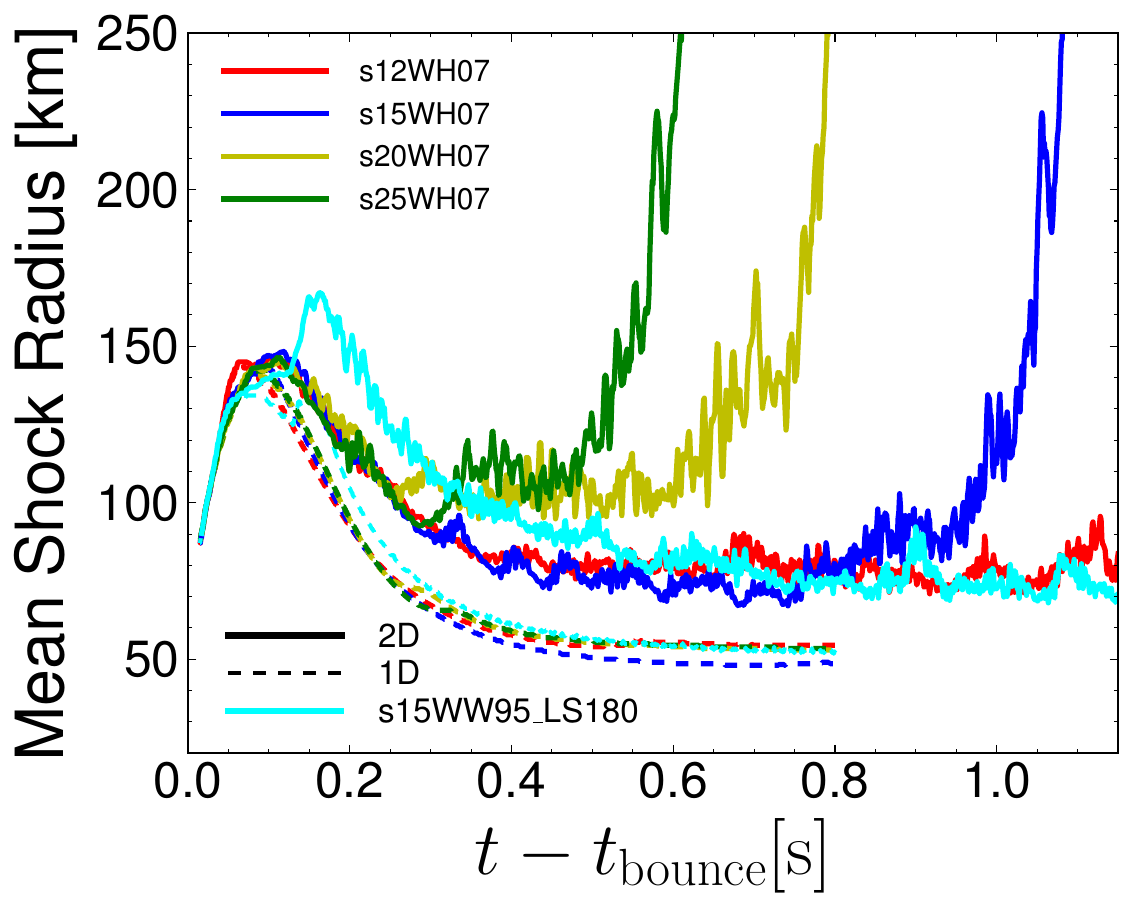}\\
\includegraphics[width=0.94\columnwidth]{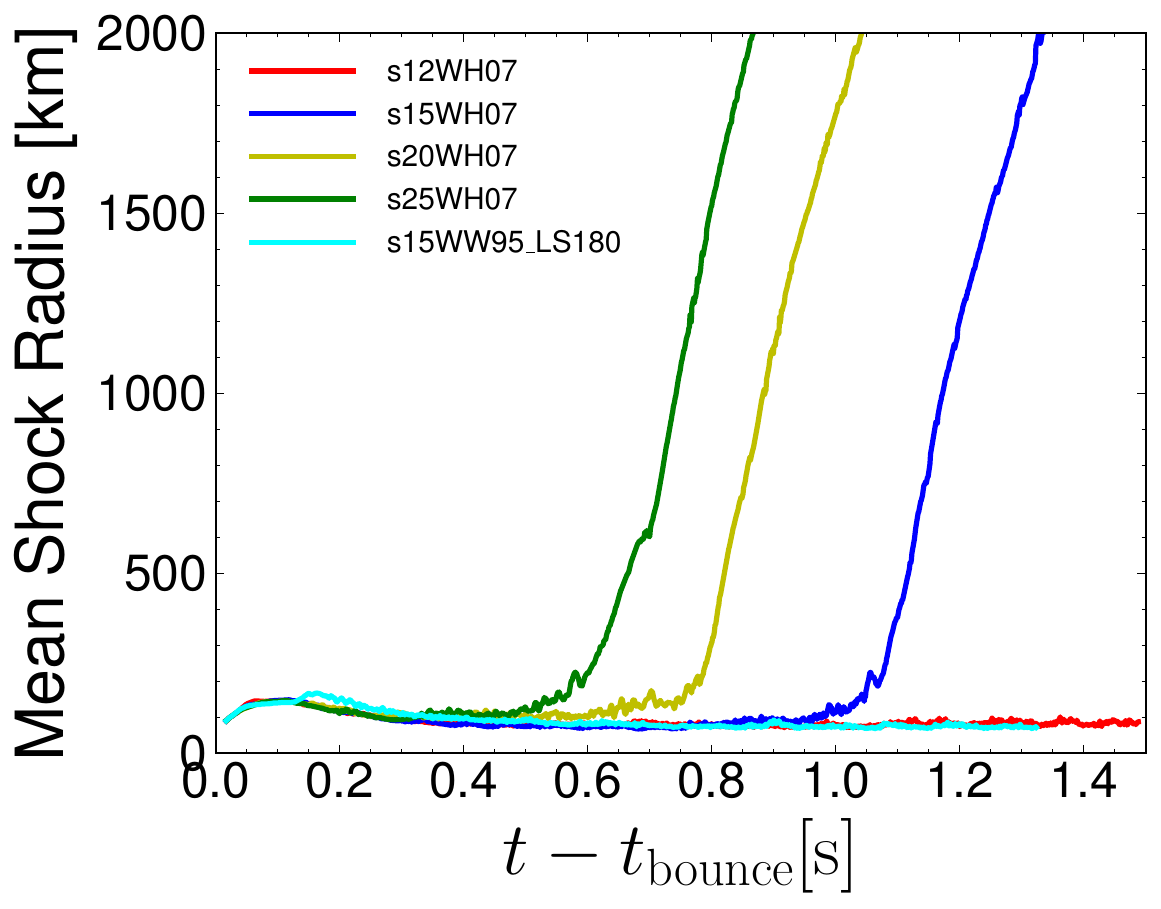}\\
\includegraphics[width=0.94\columnwidth]{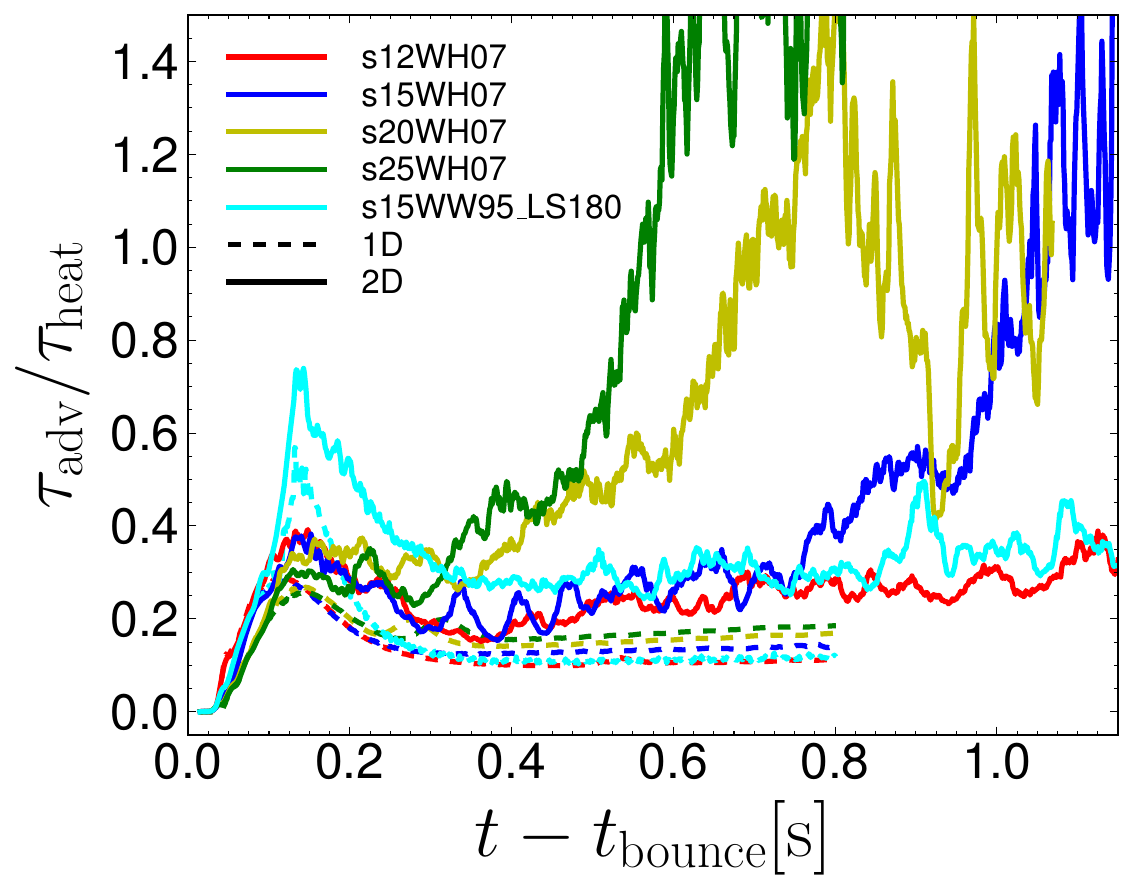}
\caption{Simulation results from 1D and 2D core collapse using GR
  effective gravity in five models. We show the time evolution of the
  mean shock radius (top and middle) and
  $\tau_\mathrm{adv}/\tau_\mathrm{heat}$ (bottom). Solid lines are
  from 2D simulations, while dashed lines are from 1D simulations.
  Models s12, s15, s20, and s25 are performed with the LS220 EOS,
  model s15WW95 uses the LS180 EOS for comparison purposes. To reduce
  the scatter, we average $\tau_\mathrm{adv}/\tau_\mathrm{heat}$ over
  a 5\,ms window. }\label{fig:GRshocks}
\end{figure}

\begin{figure*}[htbp]
\centering
\includegraphics[width=\columnwidth]{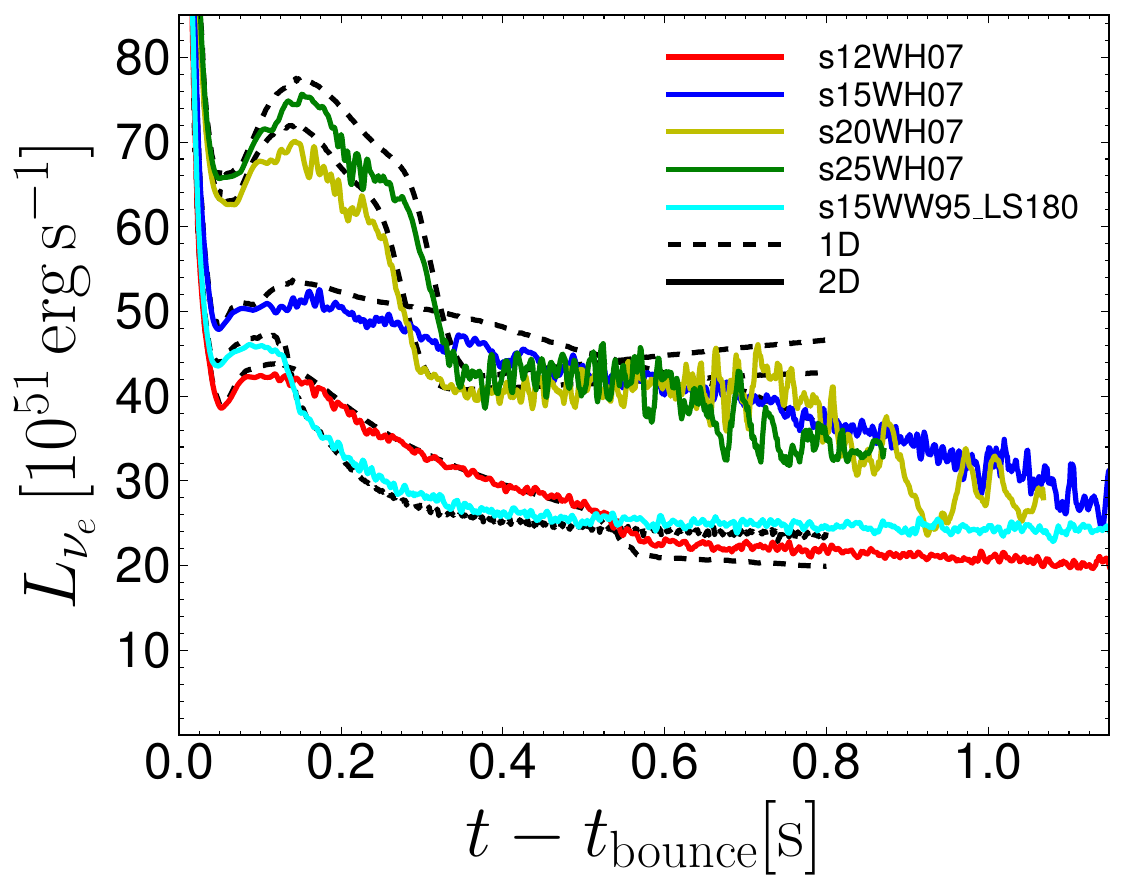}
\includegraphics[width=\columnwidth]{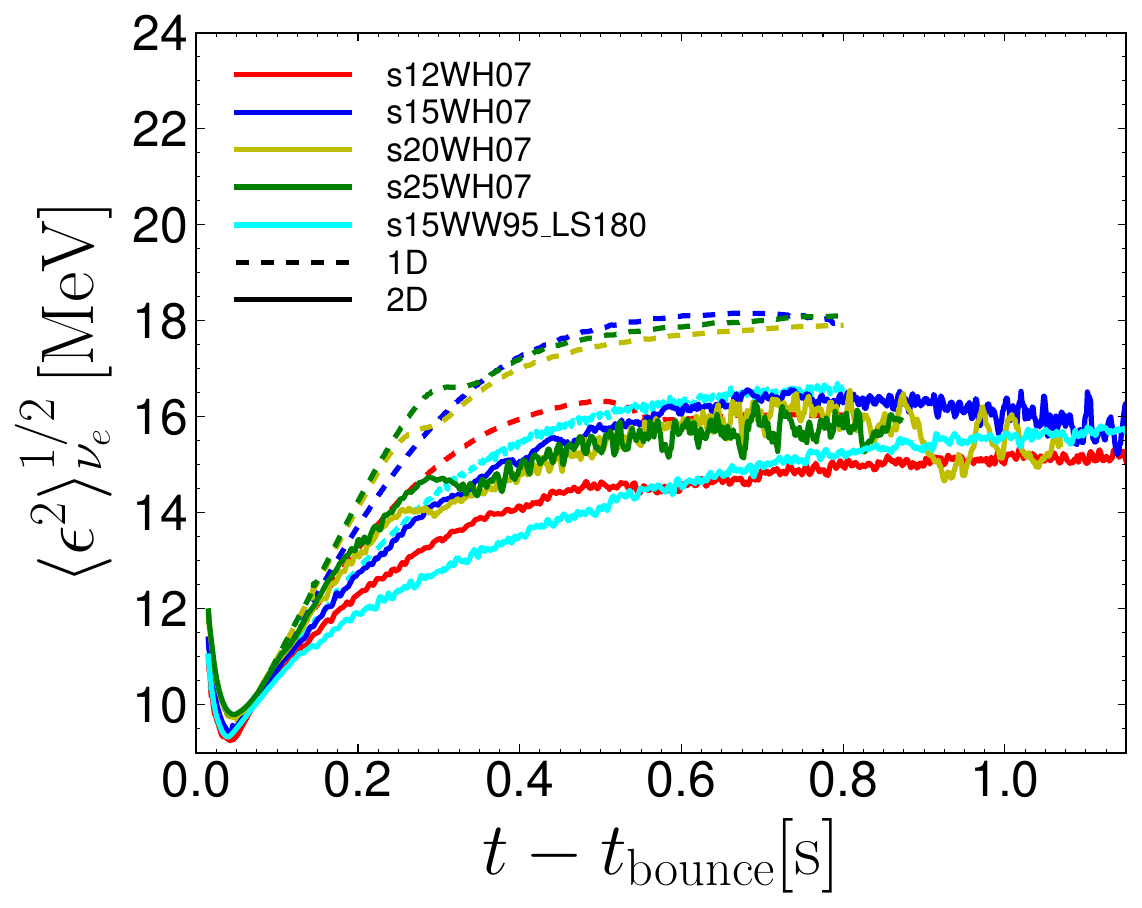}\\
\includegraphics[width=\columnwidth]{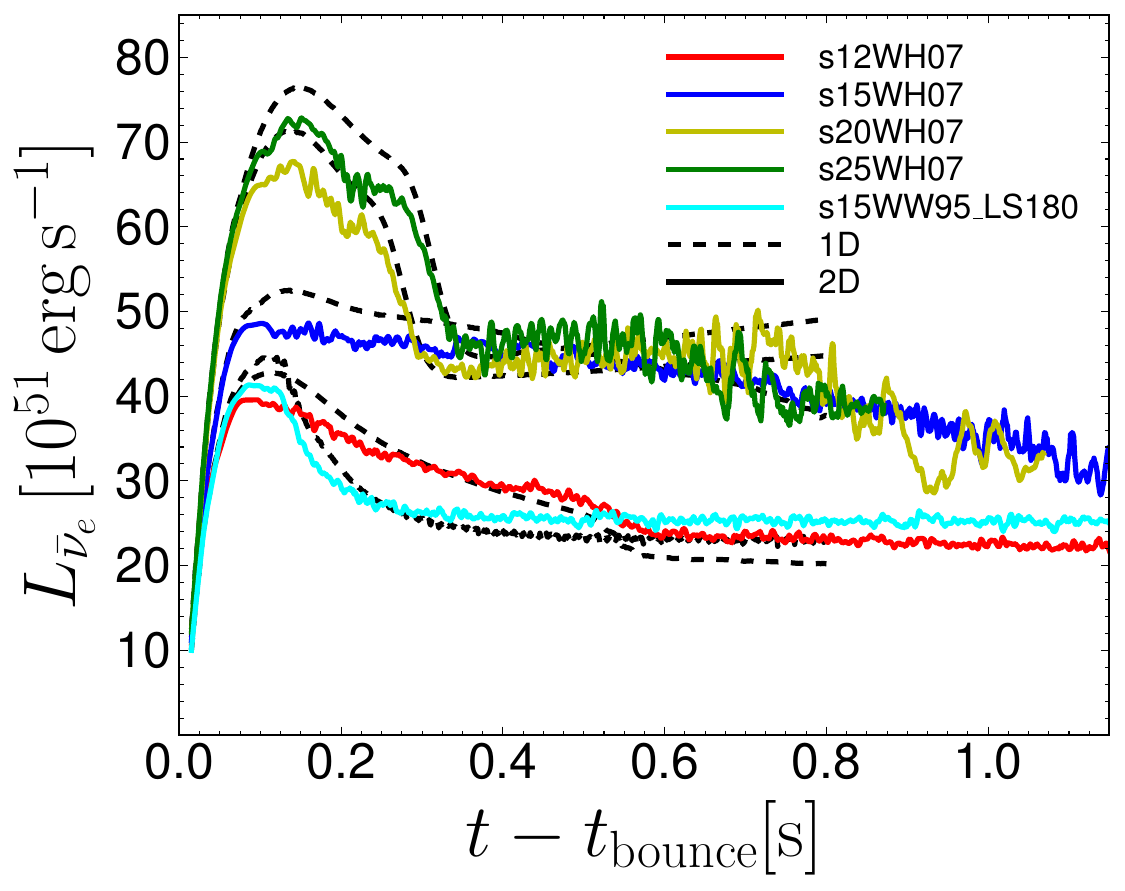}
\includegraphics[width=\columnwidth]{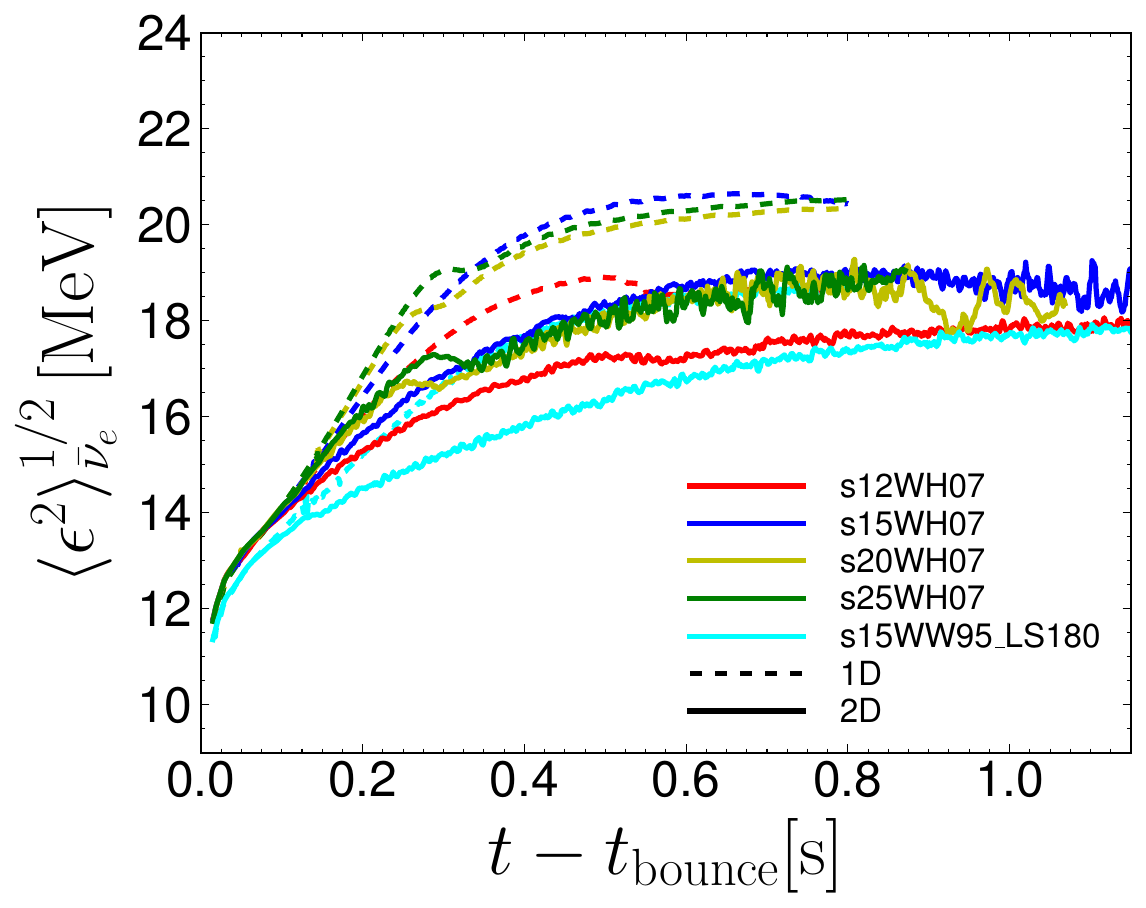}\\
\includegraphics[width=\columnwidth]{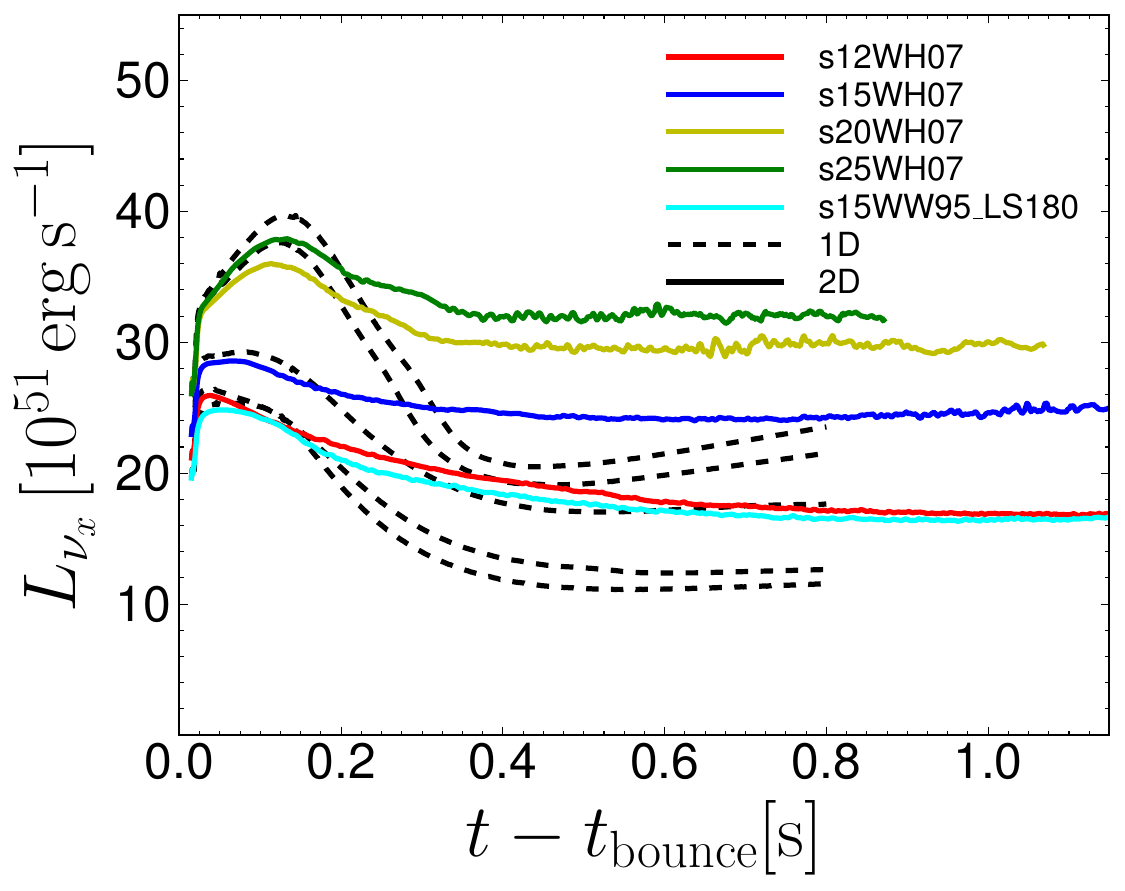}
\includegraphics[width=\columnwidth]{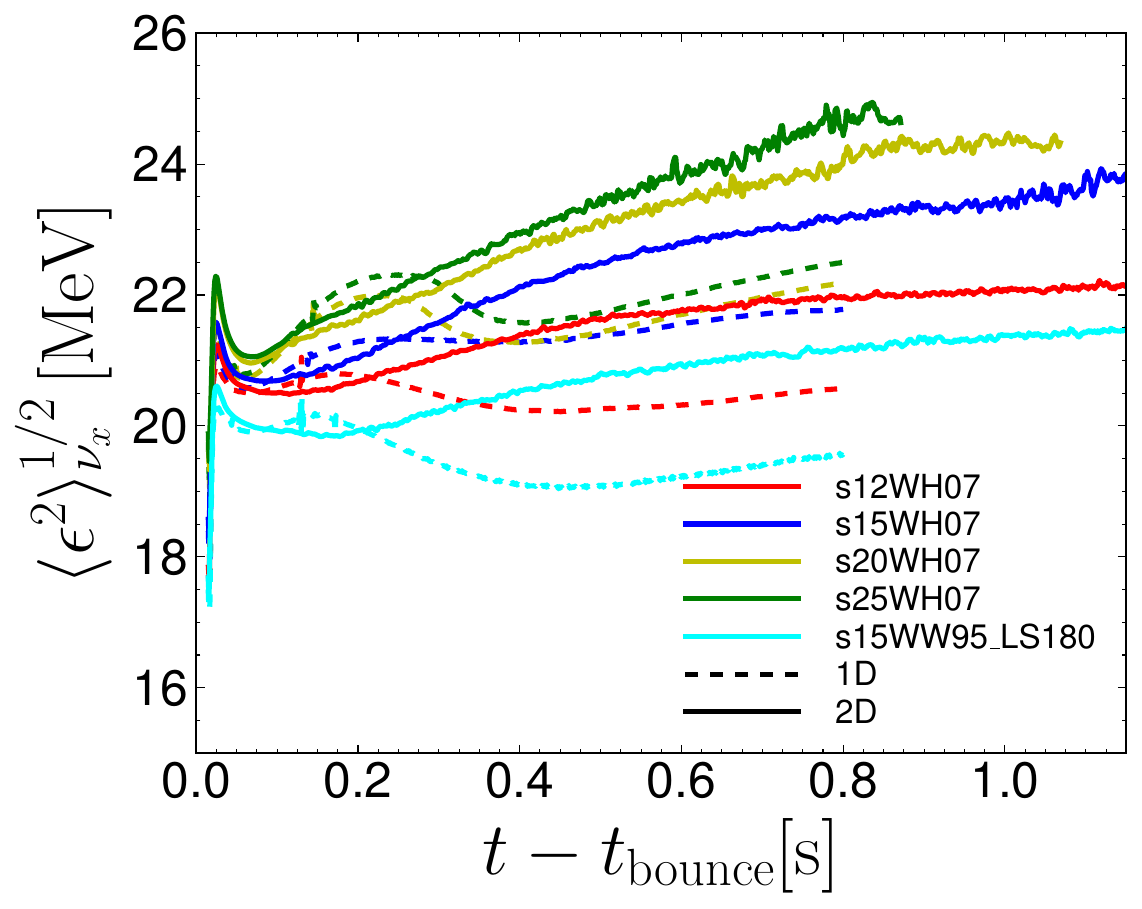}
\caption{Neutrino quantity results from 1D and 2D core collapse using
  GR effective gravity in five models. We show the time evolution of
  the $\nu_e$ (top), $\bar{\nu}_e$ (middle), and $\nu_x$ (bottom)
  luminosity (left) and root mean squared energy (right). Solid lines are from 2D simulations, while
  dashed lines are from 1D simulations.  Models s12, s15, s20, and s25
  are performed with the LS220 EOS, model s15WW95 uses the LS180 EOS
  for comparison purposes.}\label{fig:GRnus}
\end{figure*}

\subsection{2D Effective GR Results}
\label{sec:GREPresults}

In our simulations using the GR effective potential we achieve runaway
supernova shocks in our core collapse simulations for three of the
four progenitors models: s15, s20, and s25.  These explosions occur at
significantly earlier times compared to the Newtonian simulations.  We
attribute this to the increased heating in the GREP simulations
discussed in the context of 1D models presented in \sref{sec:s15WW95}.
This increased heating is a result of the more compact PNS configurations afforded by GR gravity. This gives higher electron
neutrino and antineutrino luminosities and root mean squared energies.
In \fref{fig:GRshocks}, we show the evolution of the mean shock radius
in our GREP simulations.  The top panel shows the early shock
development, while the middle panel shows the shock evolution out
to 2000\,km. The WH07 simulations start very similar to the pure
Newtonian simulations.  Initially, the shock radius extends out to
$\sim$145\,km at $\sim$100\,ms, roughly 15\,km lower than the
Newtonian simulations.  As with the Newtonian simulations, the shock
begins to recede in both 1D and 2D. By the end of the simulated time,
we achieve explosions in three of the four models.  The times at which
the mean shock radii reach 500\,km are $\sim$1100\,ms, $\sim$820\,ms,
and $\sim$670\,ms for models s15, s20, and s25, respectively.  These
values are listed in \tref{tab:literature}.  The s15WW95\_LS180 model
is markedly different than the WH07 models, the shock radius stagnates
at a slightly lower radius, $\sim$140\,km, then extends to a much larger radius
($\sim$170\,km) after the compositional interface accretes through the
shock at $\sim$150\,ms before ultimately receding and failing to be
reenergized.

In the bottom panel of \fref{fig:GRshocks}, we show the ratio of the advection time
through the gain region to the time it takes to unbind the matter in
the gain region due to neutrino heating,
$\tau_\mathrm{adv}/\tau_\mathrm{heat}$, defined in
\eref{eq:tadvovertheat}.  In models s15, s20, and s25,
$\tau_\mathrm{adv}/\tau_\mathrm{heat}$ surpasses unity at
$\sim$1\,s, $\sim$700\,ms, and $\sim$560\,ms, which corresponds to
the time when the mean shock radius is starting its outward
propagation. The s12 and s15WW95\_LS180 models never surpass the value
of unity, these models do not explode.  For the exploding models,
large, late-time asymmetries in the explosion morphology lead to values
of $\tau_\mathrm{adv}/\tau_\mathrm{heat}$ less than unity.  This is
due to the break down of the underlying assumptions of
\eref{eq:tadvovertheat}, which is technically only valid for spherical
flows.

The left column of \fref{fig:GRnus} shows the luminosity of each of the
three neutrino species.  Prior to explosion, or for the entire
evolution for models s12 and s15WW95\_LS180, the neutrino luminosities
closely follow the 1D simulation.  This is because the neutrino
luminosity being emitted in simulations with stalled or failing
supernova shocks is set by the accretion rate, which is the same in
both 1D and 2D.  The situation changes after the explosion sets in.
After this time, not all accreted matter is able to settle into the
cooling region since it is being heated by neutrinos and unbound.  As
a result, the released gravitational binding energy is not being
converted into neutrino luminosity.  The consequence is that the
post-explosion neutrino luminosity in 2D is lower than the neutrino
luminosity determined in the 1D simulation at the same time.  Our
$\nu_x$ luminosities are not as dramatically influenced by the fluid
motions in the gain region as they are emitted from deeper in the
cooling region away from the convective motions. Consequently, they
show less variability than both $\nu_e$ and $\bar{\nu}_e$. However,
there is a significant multidimensional effect on the $\nu_x$
neutrinos. We see a boost in the $\nu_x$ luminosities due to PNS
convection.  This convection slows the recession of the
neutrinosphere, and dredges up heat from deeper in the PNS.  This
results in an increase in the $\nu_x$ luminosity and $\nu_x$ root mean
squared energy (right column of \fref{fig:GRnus}) starting around
$\sim$200\,ms.  It is also responsible for lowering the electron type
neutrino and antineutrino root mean square energies.  We give a more
detailed discussion of PNS convection in \sref{sec:discussion}.

\begin{figure}[ht]
\centering
\includegraphics[width=\columnwidth]{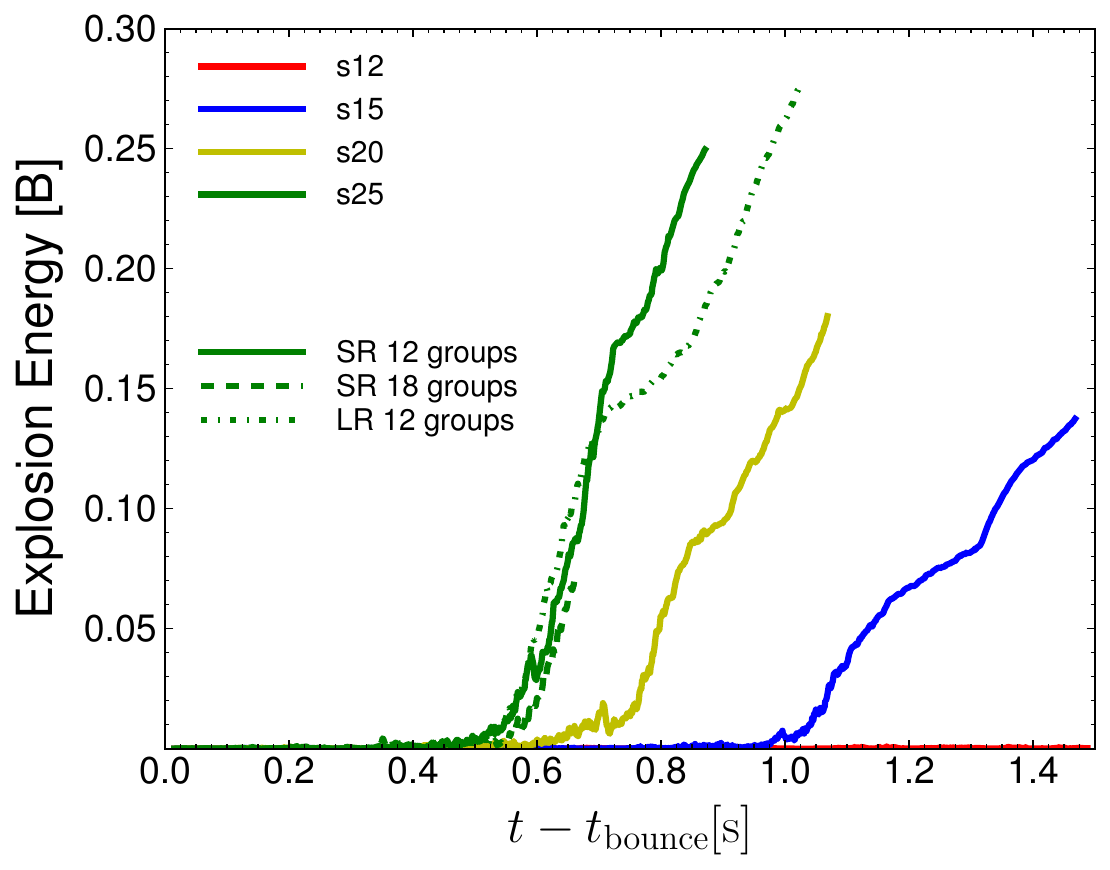}
\includegraphics[width=\columnwidth]{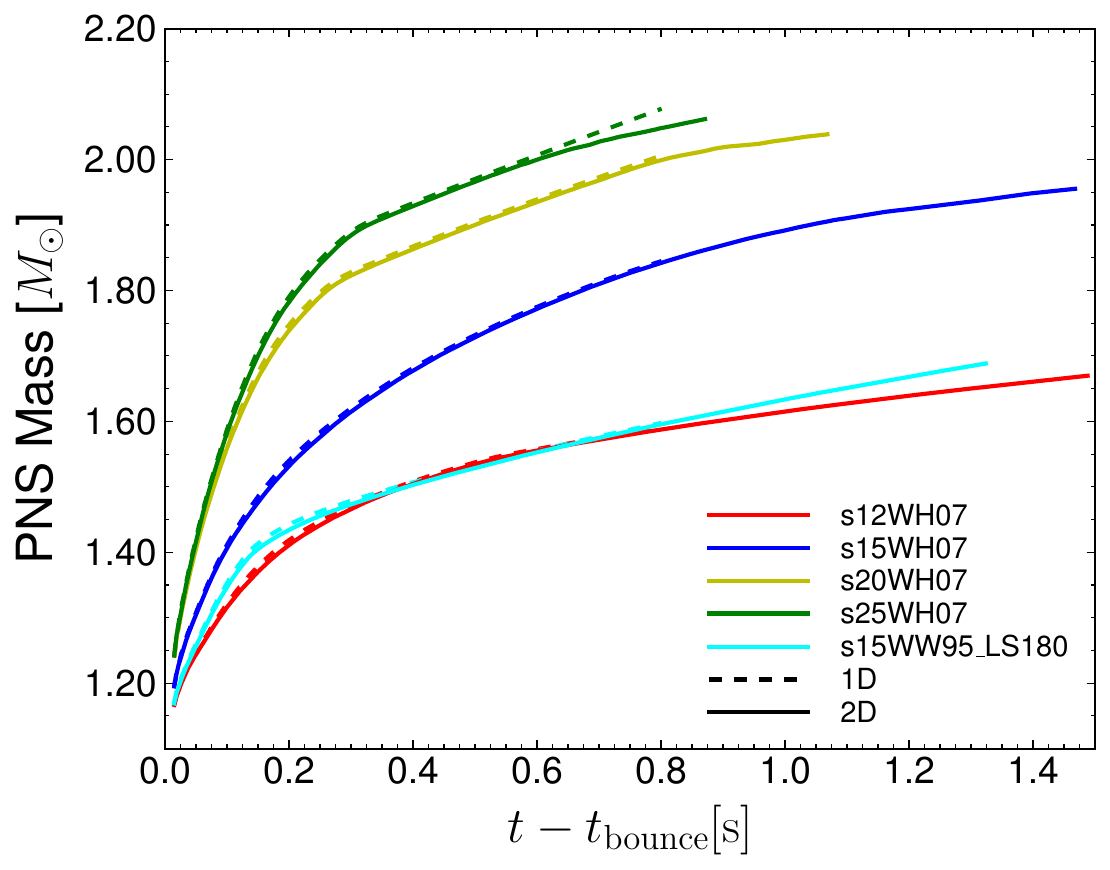}
\caption{Diagnostic explosion energies for exploding models determined
  via \eref{eq:explosionenergy} (top panel) and cumulative PNS
  baryonic mass (bottom panel). The explosion energy estimate includes
  an estimate of the nuclear recombination energy, but does not
  include the binding energy of the overlying star. We also show
  (dashed and dashed-dotted lines) tests with the s25 model where we
  decrease the resolution and where we decreased the resolution but
  increase the number of neutrino energy bins. For the bottom panel,
  the PNS is defined as the region with matter densities larger than
  $10^{11}$\,g\,cm$^{-3}$.  Solid lines denote the mass from 2D
  simulations while dashed lines show the PNS mass from 1D
  simulations.}\label{fig:explosionenergy}
\end{figure}

\begin{figure*}[ht]
\centering
\includegraphics[trim=0.0cm 4.0cm 1.5cm 14cm, angle=270, clip,  width=0.66\columnwidth]{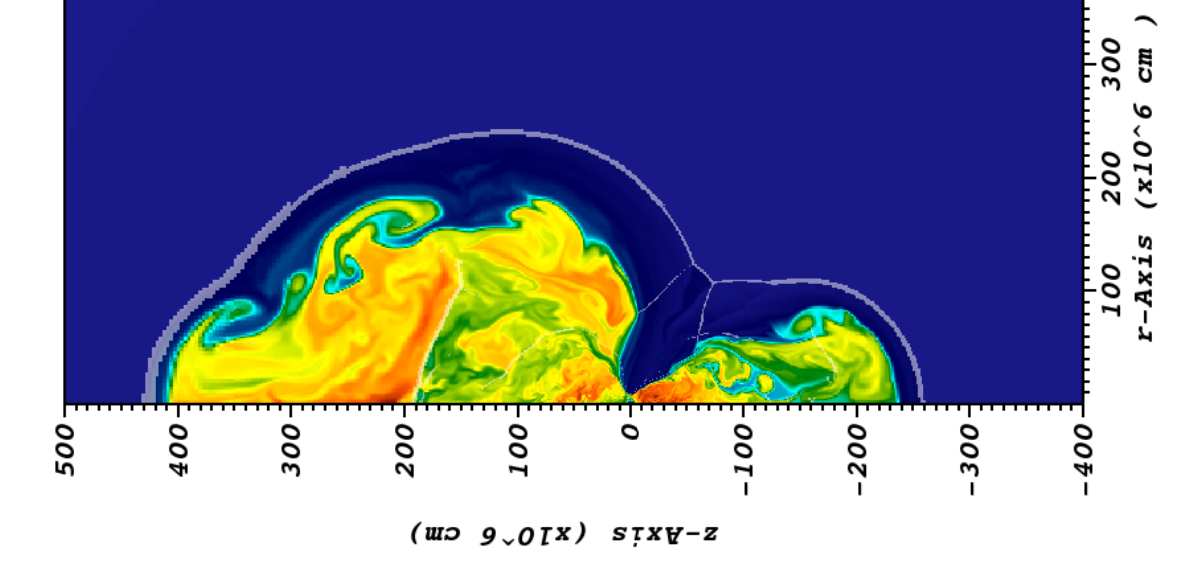}
\includegraphics[trim=0.0cm 4.0cm 1.5cm 14cm, angle=270, clip,  width=0.66\columnwidth]{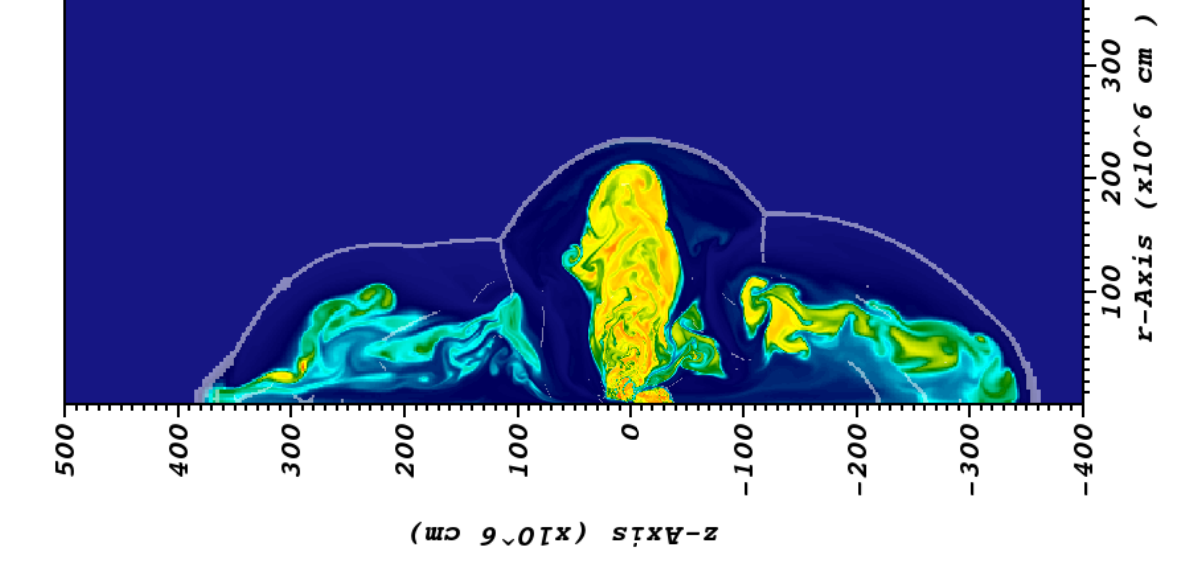}
\includegraphics[trim=0.0cm 4.0cm 1.5cm 14cm, angle=270, clip,  width=0.66\columnwidth]{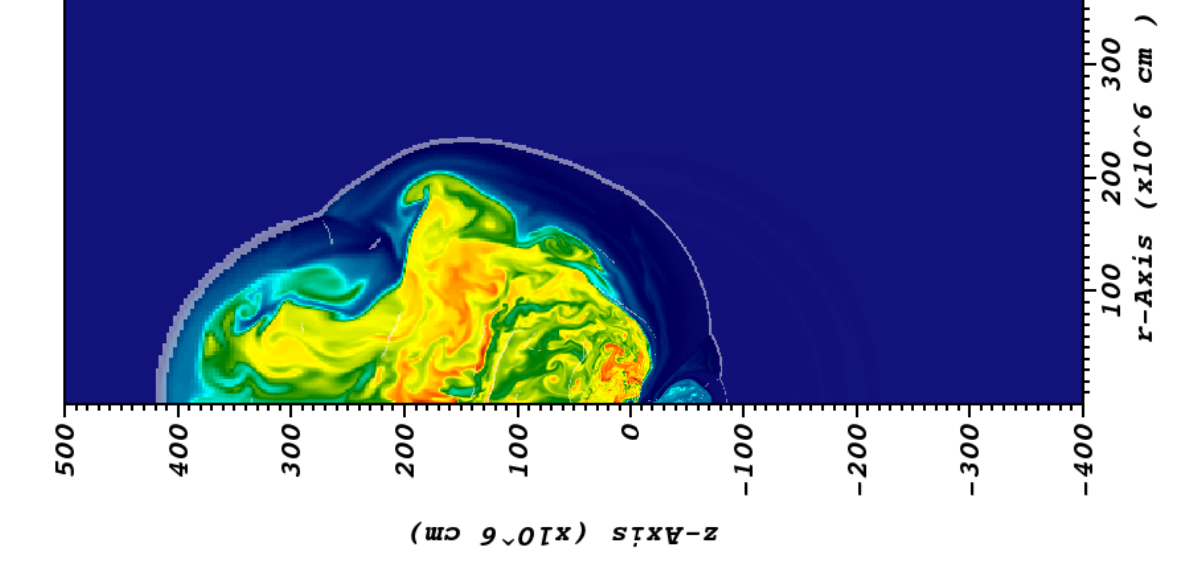}
\caption{Entropy color plots of exploding models s15, s20, and s25 at
  1340\,ms, 1070\,ms, and 830\,ms after bounce, respectively.  These
  times correspond to when the shock front first reaches 4000\,km.
  The explosions are strongly asymmetric, indicative of weak explosions.
  Grey zones correspond to the shock
  position and blue $\to$ green $\to$ yellow $\to$ orange $\to$ red
  colors denote increasing values of the entropy}\label{fig:4000kmsnapshots}
\end{figure*}

In the top panel of \fref{fig:explosionenergy}, we show the early development of the
diagnostic explosion energy \citep{buras:06a,bruenn:13,melson:15b},
\begin{equation}
E_\mathrm{exp}^\mathrm{diag} = \int_\mathcal{V} d\mathcal{V} \rho [ 1/2|v|^2 +
(\epsilon-\epsilon_0) + \phi ]_{> 0}\,,\label{eq:explosionenergy}
\end{equation}
where the volume integral is over all zones where the integrand is
positive.  $\phi$ is the gravitational potential energy, $\epsilon$ is
the specific internal energy and $\epsilon_0$ is the specific internal
energy of matter with the same density and $Y_e$, but $T=0$.  This
roughly accounts for the energy that will be released into kinetic
energy as the matter cools and the free nucleons recombine into alpha
particles, and later nuclei. To truly determine this energy, one must
track the evolution to much longer times and do a proper treatment of
nucleosynthesis, something we are not capable of doing with our
current infrastructure.  We also note that this diagnostic explosion
energy does not account for the binding energy of the material outside
the shock (i.e, the ``overburden'') which can be significant in the
higher mass models, perhaps upwards of 1\,B \citep{bruenn:16}.  Our
explosion energies are quite small compared to the results of
\cite{bruenn:13}, and more along the lines predicted from other
successful explosions in two and three dimensions \citep{hanke:12,
  melson:15b, mueller:12a, suwa:10, pan:16, suwa:16}. We do note that
the amount of simulated post-explosion time is short.  In our
simulations, along with the weak explosions comes continued accretion
after the explosion sets in. We show this in the bottom panel of
\fref{fig:explosionenergy}, where we show the PNS mass (PNS defined as
the region with a matter density of at least $10^{11}$\,g\,cm$^{-3}$)
in 2D (solid lines) and 1D (dashed lines).  The simulations which do
not produce explosions have similar PNS masses in 1D and 2D.  For the
simulations that explode, the growth of the PNS mass slows after the
explosion sets in, but is not zero in any model at the latest times
simulated.  Finally, we note that the explosions are highly
non-spherical with strong accretion occurring along various
directions.  We show graphical snapshots of our three exploding models
in \fref{fig:4000kmsnapshots} at a time determined by when the maximum
shock radius reaches 4000\,km, corresponding to 1340\,ms, 1070\,ms,
and 830\,ms after bounce for models s15, s20, and s25, respectively.

\subsubsection{Resolution and Random Perturbations}

We simulate core collapse in model s25 with a lower resolution.  As
discussed in \S~\ref{sec:simsetup}, our production simulations have a
central cell size of $\sim$488\,m and maintain an effective angular
($\Delta r /r $) resolution from at least 0.5 degrees outside
$\sim$55\,km. This results in the first refinement decrement (from
$\sim$500m to $\sim$1\,km) at $\sim$107\,km. For our lower resolution
simulation, we only enforce an effective resolution of 0.9 degrees,
giving refinement decrements at $\sim$65\,km, $\sim$139\,km, and so
on. In addition to this simulation we also simulate model s25 at our
standard resolution and with 18 energy groups instead of our standard
12. In both tests, explosions develop at similar times to our baseline
model. The development of the explosion energies for the successful
models (see dashed and dashed-dotted lines in
\fref{fig:explosionenergy}) are similar between these tests.

\begin{figure}[ht]
\centering
\includegraphics[width=\columnwidth]{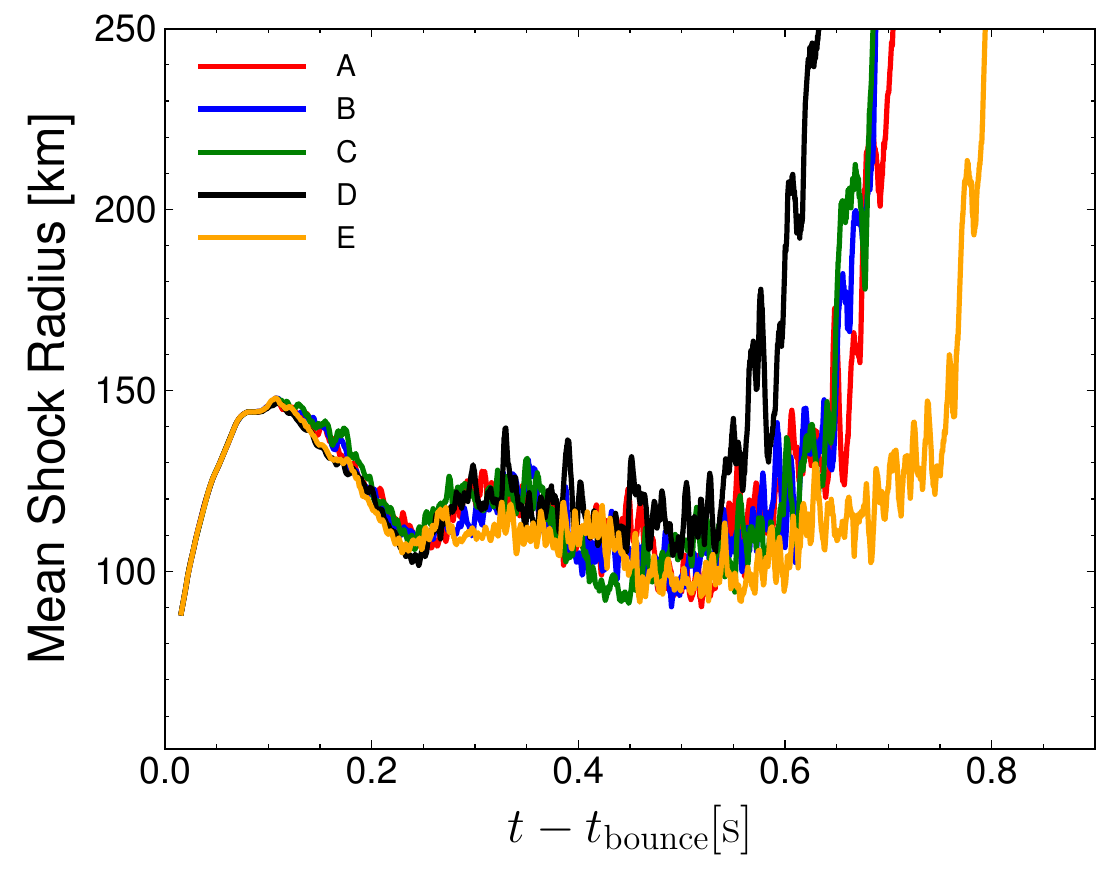}
\caption{Shock radius evolution for five simulations of model s20, each
  started with different random density perturbations.  The explosion
  times differ due to stochastic variations in the postshock
  region.}\label{fig:s20random}
\end{figure}

As a further test, we explore the effect of imposed random
perturbations on the development of the explosion.  We simulate model
s20 five times, each time perturbing the initial density field
(when we transition from our 1D model to the 2D simulations at $\sim$15\,ms
after bounce) randomly (and uniformly) by up to $\pm 0.1\%$. We show
the evolution of the shock radii in \fref{fig:s20random}.  Stochastic
variations in the hydrodynamic instabilities behind the shock cause
the explosion times (determined by when the mean shock passes 400km)
to vary from $\sim$670\,ms to $\sim$830\,ms.  The average is $\sim$723\,ms with a
formal variance of $\sim$63\,ms.  We note that the explosion time of our
main simulation of s20 falls near the latter end of this range. These
simulations are also discussed in \cite{oconnor:17}.

\section{Discussion}
\label{sec:discussion}
\subsection{GR vs Newtonian Gravity}

\begin{figure}[ht]
\centering
\includegraphics[width=0.94\columnwidth]{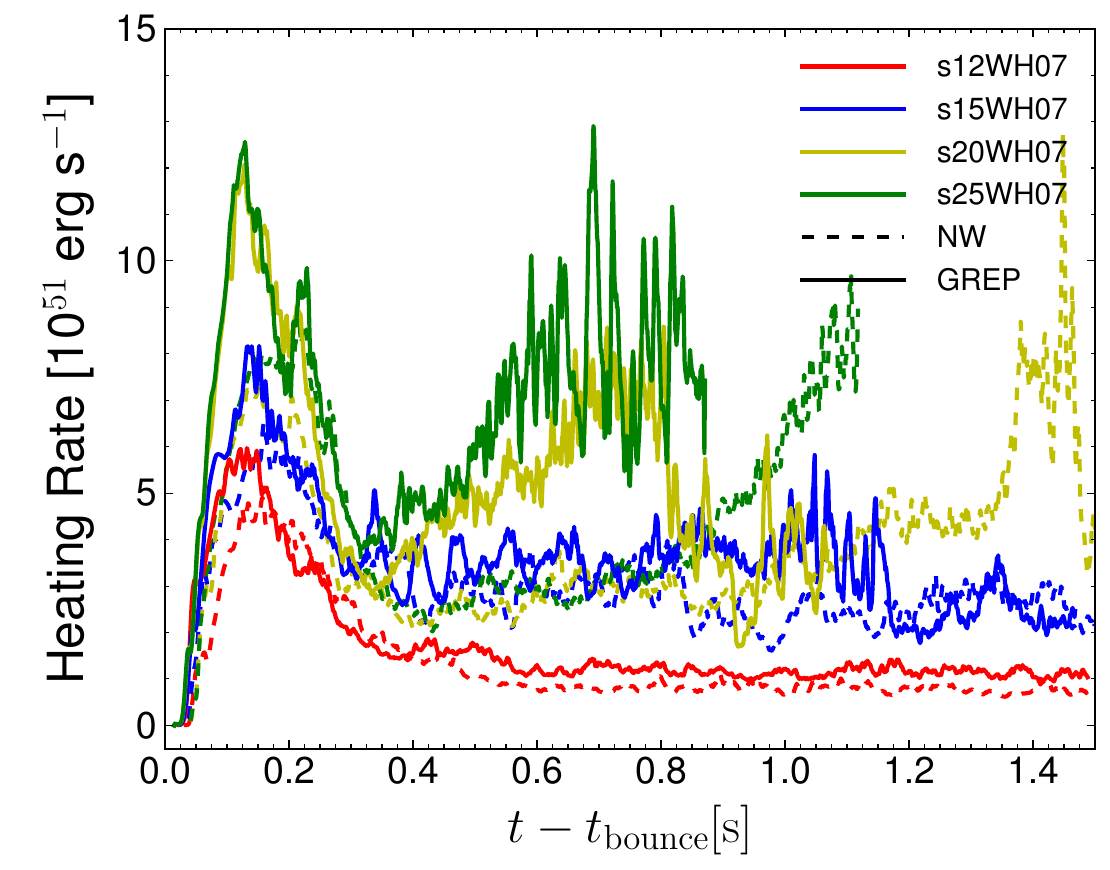}\\
\includegraphics[width=0.94\columnwidth]{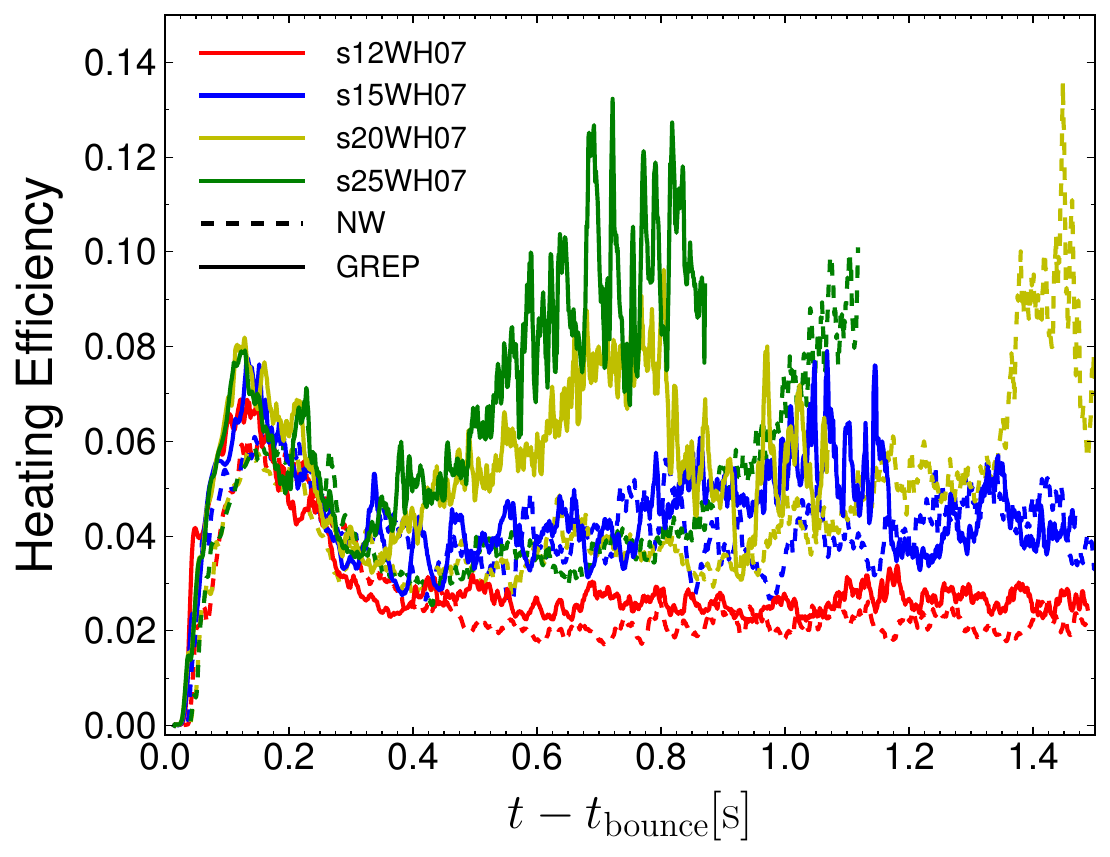}\\
\includegraphics[width=0.94\columnwidth]{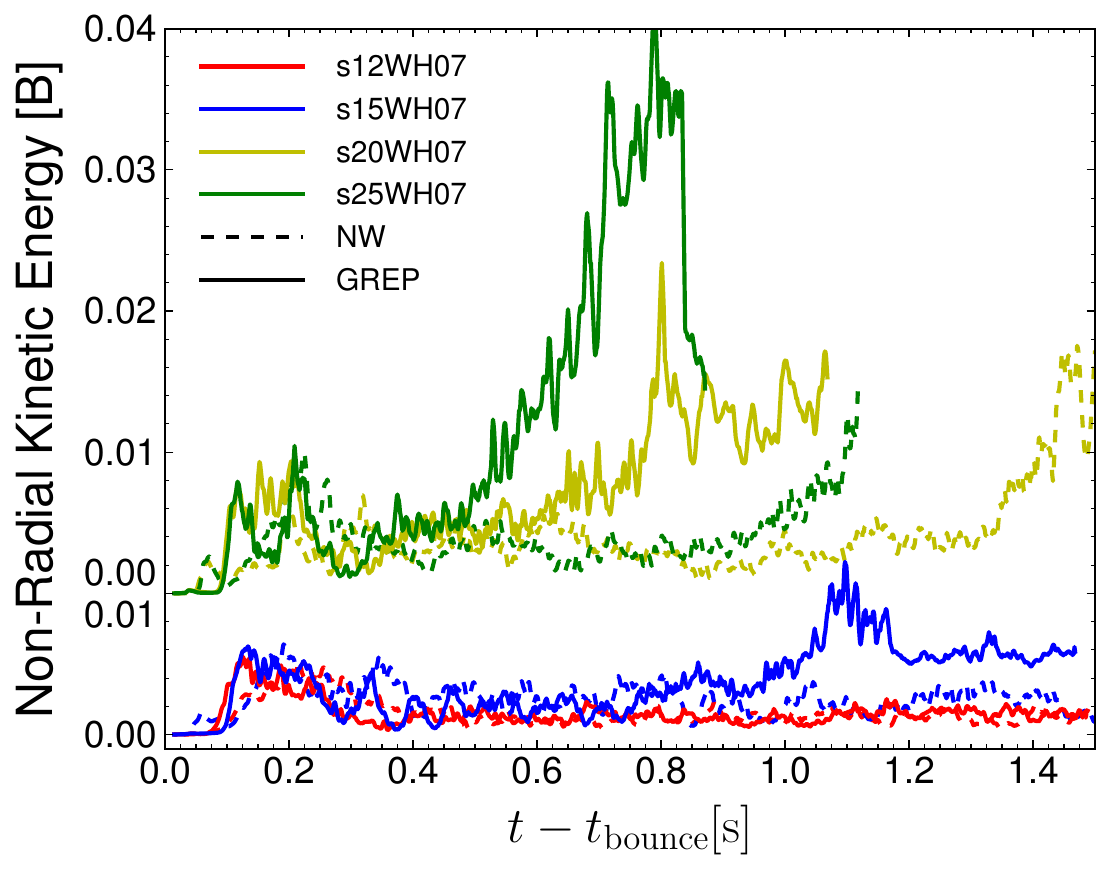}\\
\caption{Neutrino heating metrics for 2D simulations with both GR
  effective potential gravity and Newtonian gravity.  We show the net
  heating (top panel), the heating efficiency (middle panel), and
  non-radial kinetic energies in the gain region (bottom panel) for
  each of the four WH07 models and model s15WW95\_LS180 used in this
  study.  Simulations with Newtonian gravity are shown as dashed lines
  while the GR effective potential simulations are solid lines. For
  the bottom panel, models s20 and s25 are shifted up for clarity. All
  curves are averaged over 5ms to reduce
  noise.}\label{fig:GREPvsNW_2D}
\end{figure}

From the results presented in the last section, it is clear that the
approximation of Newtonian gravity gives less favorable conditions for
a successful core-collapse supernova explosion than the more realistic
GR treatment.  Even though we do not include full GR gravity and
hydrodynamics, the test results of \sref{sec:s15WW95}, and the
comparisons to the previous work of
\cite{liebendoerfer:05,marek:06,oconnor:15a} convince us that using an
effective GR potential allows us to capture the important aspects of
GR for modeling core-collapse supernovae.  Most importantly, the 1D
results of \sref{sec:s15WW95} show that GR gravity leads to higher
neutrino luminosities, and higher root mean squared energies which
directly follow from a more compact PNS structure.  The neutrino
mechanism relies on the capture of neutrinos emitted from the cooling
region into the gain region.  This capture is greater with higher
luminosities, and more efficient with higher neutrino energies.
However, in 1D, the added benefits of GR do not lead to successful
explosions as has been previously demonstrated
\citep{liebendoerfer:01b}.  This trend of increased heating and higher
heating efficiencies carries over to 2D, which we demonstrate in
\fref{fig:GREPvsNW_2D}. In this figure, we show the net heating rate
in the gain region (top), the heating efficiency (middle;
\eref{eq:etaheat}), and the non-radial kinetic energy in the gain
region (bottom) for the four WH07 models using the GR effective
potential gravity (solid lines) and Newtonian gravity (dashed
lines). In all models, both the heating rate and the heating
efficiency are larger in the simulations using the GR effective
potential, by up to $\sim$20-30\% between 100\,ms and 200\,ms after
bounce. Furthermore, the appearance of the gain region occurs sooner
for the GREP models by $\sim$20\,ms. We also generally see more
non-radial kinetic energy in the gain region, which is a sign of
stronger neutrino-driven convection. As a consequence of these
improved neutrino heating conditions, we see earlier explosions in the
GREP models s20 and s25 by $\sim$0.5\,s and an explosion in
the GREP model s15 where the Newtonian model did not explode.

\subsection{Comparison to Literature}
\subsubsection{Newtonian Gravity}

In \tref{tab:literature} we have summarized other studies in the
literature which simulate the four models simulated here and use
Newtonian gravity. They include \cite{dolence:15, suwa:16, pan:16}.
Many aspects of our results compare quite well to the Newtonian
results of \cite{dolence:15}. \cite{dolence:15} use the same
progenitor models, the same neutrino microphysics (based on
\citealt{bruenn:85,brt:06}), and also a truly multidimensional neutrino
transport scheme, albeit MGFLD. Matching our simulations, none of the
2D simulations in \cite{dolence:15} explode before 600\,ms, which is
when their simulations stop.  Comparison of
$\tau_\mathrm{adv}/\tau_\mathrm{heat}$ show good agreement. At
$\sim$200\,ms, $\tau_\mathrm{adv}/\tau_\mathrm{heat} \sim 0.3-0.4$ in
both simulation sets and for all four models. Furthermore, our heating
efficiencies (see \fref{fig:GREPvsNW_2D}) near $\sim$200\,ms after
bounce are $\sim$0.06, similar to the values of
$\sim$0.065-$\sim$0.085 seen by \cite{dolence:15}. Two-dimensional
Newtonian simulations of these models are also presented in
\cite{suwa:16} using IDSA for neutrino transport.  There, the LS220
EOS is used.  They observe an early explosion in a few models,
including s12, and failure in the majority of the rest. One issue
preventing a detailed comparison is that \cite{suwa:16} ignore
heavy-lepton neutrinos in their simulations.  It is likely the case
that this prevents the PNS from cooling and causes the shock to
initially stall at $\sim240$\,km and hover around $\sim$200\,km in
failed models.  \cite{pan:16} also perform Newtonian axisymmetric
simulations of core collapse using IDSA for neutrino
transport. However, they find early explosions for all progenitors
simulated: 11 \msun, 15 \msun, 21 \msun, and 27 \msun~models from
\citet{whw:02} as well as two of the models explored here: s15 and
s20. It is worth noting that later simulations in \cite{pan:17}
find no explosions in models s15 and s20 with Newtonian gravity.

\subsubsection{GR Gravity}

We also compare to studies in the literature that use GREP
gravity. These include \cite{summa:16, bruenn:16}.  We
will first discuss the results of \cite{summa:16}. \cite{summa:16}
achieve explosions in all four models studied here.  Their explosion
times are $\sim$790\,ms, $\sim$620\,ms, $\sim$320\,ms and
$\sim$400\,ms for models s12, s15, s20, and s25, respectively.  These
explosions times are earlier than ours by $\sim$300-$\sim$500\,ms.  We
do not see an explosion in model s12 before $\sim$1.5\,s, when our
simulation was stopped. The evolution of the mean shock radius is
qualitatively similar. The initial stalling radius in our simulations
is $\sim$150\,km compared to $\sim$160\,km in \cite{summa:16}. After
this, there is a period of shock recession before the ultimate shock
expansion. However, the simulations of \cite{summa:16} have a much
more pronounced shock recession. This may stem from the differences in
the neutrino interactions used. \cite{summa:16} use inelastic
neutrino-nucleon scattering and absorption opacities from
\cite{burrows:98,burrows:99} whereas we use simpler elastic rates from
\cite{bruenn:85}. They use microphysical electron capture rates on
heavy nuclei for the collapse phase instead of the approximate form
from \cite{bruenn:85} implemented here.  Finally, \cite{summa:16}
include neutrino-neutrino pair scattering and annihilation (to other
neutrino pairs) as additional pair-processes. One important
consequence of these improved opacities is the more efficient cooling
of the PNS which gives higher $\nu_x$ luminosities and faster
contraction and as a consequence. In \cite{oconnor:17}, we have made
the first steps to improve our neutrino interactions by including a
correction factor of the neutral-current scattering cross sections
that mimics \cite{burrows:98,burrows:99}.  There we found earlier
explosion times by $\sim$100-150\,ms. Given these lingering
differences between our simulations and those of \cite{summa:16}, a
comparison with even closer underlying physics is warranted. Although
we find it quite encouraging that we find relatively close agreement
on many quantities, including neutrino luminosities, neutrino
energies, neutrino heating rates and efficiencies, and explosion
diagnostics like $\tau_\mathrm{adv}/\tau_\mathrm{heat}$, even with the
differences between our simulation codes and simulation techniques.

In contrast, our results are in qualitative disagreement with the
results of \cite{bruenn:13,bruenn:16} (labeled here as Bruenn \emph{et
  al.}).  At 100\,ms, all four models in Bruenn \emph{et al.} stall at
a radius of $\sim$200\,km. However, after the shock stalls it
continues to slowly move out in radius until it ultimately begins to
run away. There is no period of shock recession in any model. We note,
however, an updated 2D simulation in \cite{lentz:15} of model s15
gives a stall radius of $\sim$175\,km, there is still no recession of
the shock front in this updated simulation.  We can also directly
compare neutrino quantities and heating rates with the results in
Bruenn \emph{et al}.  The neutrino luminosities predicted from both
sets of simulations agree within $\sim$5\% (this difference is in part
due to the fact that Bruenn \emph{et al.} measure their quantities in
the co-moving frame at 1000\,km where the fluid velocities are
$\sim0.03\,c$). The root mean squared energies (right column of
\fref{fig:GRnus}) show a larger difference, Bruenn \emph{et al.}
predict neutrino energies up to $\sim20\%$ larger than our results at
$\sim$100\,ms after bounce, this difference cannot be explained by the
difference between the co-moving frame measurements alone, but could
be due to differences in the neutrino interactions (which include some
of the improved rates included in \cite{summa:16} listed above, but
not the neutrino-neutrino scattering or annihilation to other neutrino
pairs) or the neutrino transport (which is MGFLD, a more approximate,
one moment transport scheme using the ray-by-ray+
approximation). Roughly, the neutrino heating rates scale with both
the electron-type neutrino luminosity and the electron-type mean
\emph{squared} neutrino energy, therefore we expect to see the
influence of the higher root mean squared energies seen in Bruenn
\emph{et al.} reflected in the heating rates. Our GREP simulations
have a net heating rate in the gain region at 100\,ms after bounce of
$\sim$6\,B\,$s^{-1}$, $\sim$6\,B\,$s^{-1}$ $\sim$10\,B\,$s^{-1}$,
and $\sim$10\,B\,$s^{-1}$ for models s12, s15, s20, and s25,
respectively, while the heating rate in Bruenn \emph{et al.}  at the
same time are $\sim$9\,B\,$s^{-1}$, $\sim$13\,B\,$s^{-1}$,
$\sim$20\,B\,$s^{-1}$, and $\sim$20\,B\,$s^{-1}$, for models s12, s15,
s20, and s25, respectively (the heating rate at 100\,ms after bounce
in the updated 2D simulation of model s15 in \cite{lentz:15} is
9\,B\,s$^{-1}$).  The heating rates seen in the Bruenn \emph{et al.}
simulations are roughly $\sim$50\% higher than ours. This large
discrepancy at such an early postbounce time warrants further
investigation beyond comparison of literature results.

\subsubsection{1D Parameterized Models}

\begin{figure}[ht]
\centering
\includegraphics[width=\columnwidth]{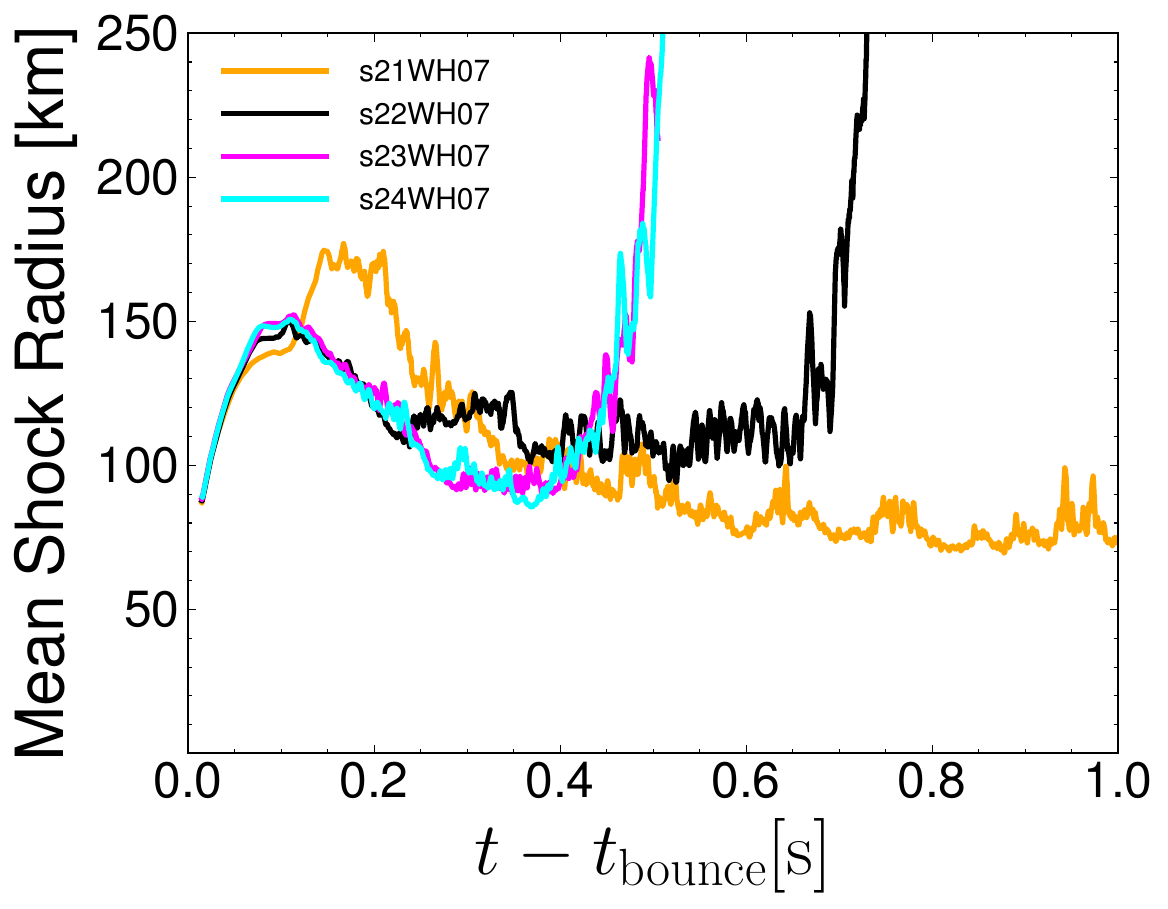}
\caption{Shock radius evolution for four simulations of models s21,
  s22, s23, and s24. All models except s21 explode,
  contrary to predictions based on 1D parameterized
  models.}\label{fig:othersWH07}
\end{figure}

We note that the heating efficiency of successful models is
significantly less than the heating efficiency one would predict was
required for an explosion in 1D. For example, \cite{oconnor:11}
predicts a critical heating efficiency of 0.172, 0.176, and 0.185 for
models s15, s20, and s25. The success of these models in our
simulations is not due solely to the increase in neutrino heating as a
result of multidimensional instabilities, rather the instabilities
themselves are directly contributing to success of the explosion
\citep{couch:15a}.  Our results, while only consisting of four models,
are inconsistent with the 1D parameterized model predictions of
success/failure from \cite{ertl:16}: our exploding models are
predicted to fail and our failed model is predicted to succeed.  To
extend this further we have simulated four additional models, s21,
s22, s23, and s24 from \cite{woosley:07}. Based on \cite{ertl:16}, s21
is predicted to explode and the remaining models are predicted to
fail. In particular, models s22 through s25 are in a region of ZAMS
mass that many studies have predicted will from failed supernovae and
black holes.  As can be seen in \fref{fig:othersWH07}, this is not the
case, similar to the pattern observed in the other models, the s21 model
fails to explode while the other three models successfully explode.
We note the similarity of the shock evolution in s21 to that of
s15WW95. These two models have remarkably similar presupernova
structure in the inner regions, characterized by a steep gradient in
the density at a compositional interface at relatively low mass
coordinate (compared to the other models explored here). The ultimate
predictive power of 1D parameterized explosion models like these must
be further tested (or calibrated) with multidimensional models,
preferably three dimensional models. The impact of the
multidimensional dynamics on the explosion mechanism is likely
significant enough that modeling it through increased neutrino heating
alone may not be sufficient.

\subsection{Impact of Protoneutron Star Convection}

\begin{figure*}[ht]
\centering
\includegraphics[trim=6.0cm 2.5cm 3.5cm 2.0cm, clip,  width=\textwidth]{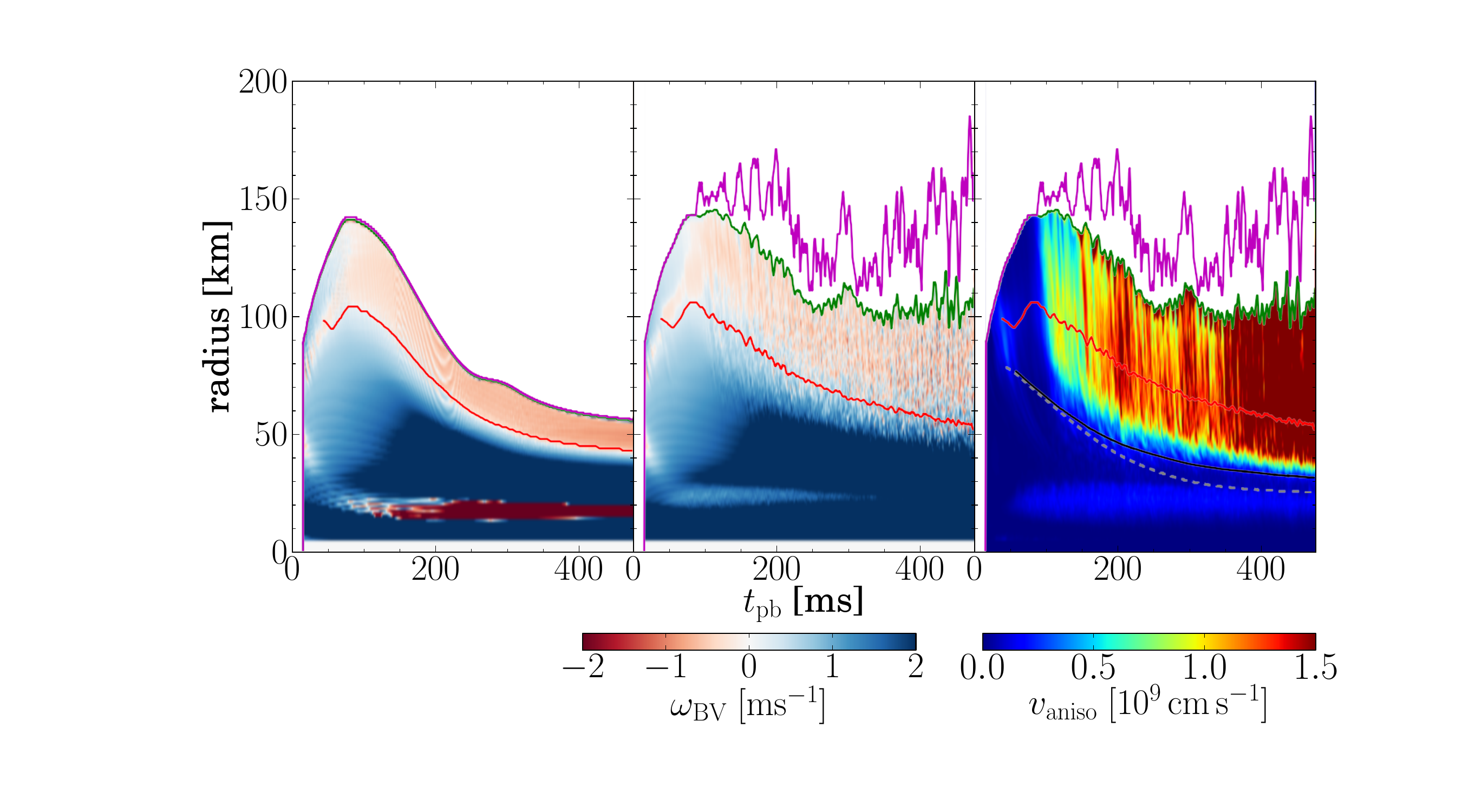}
\caption{Color maps of the Brunt-V\"ais\"al\"a frequency (left,
  center) from the 1D and 2D GREP s20 simulations as a function of
  time and radius. Red colors denote convectively unstable regions while blue
  regions are convectively stable. The right panel shows the
  anisotropic velocity of the 2D model. Shown as green, magenta, and red
  lines are the mean shock radius, maximum shock radius, and gain
  radius, respectively.  We mask out the color map above the mean
  shock radius. In the right panel, the black and grey lines show the
  2D and 1D, respectively, heavy lepton neutrinosphere.}\label{fig:s20WH07_omega}
\end{figure*}

Our 2D models develop PNS convection starting at $\sim$60\,ms after
bounce.  As mentioned earlier, this convection has a quantitative
impact on the neutrino luminosities, particularly the heavy lepton
neutrino luminosities.  We will discuss PNS convection in more detail here.

In \fref{fig:s20WH07_omega}, we show color maps of the
Brunt-V\"ais\"al\"a frequency, $\omega_\mathrm{BV}$, as a function of
time and radius for the GREP model s20, for both 1D (left) and 2D
(center). While convection will not develop in our 1D model, examining
$\omega_\mathrm{BV}$ is useful nonetheless.  For each time slice
(angle averaging the matter properties in the case of 2D) we compute
$\omega^2_\mathrm{BV}$ \citep{foglizzo:06},
\begin{equation}
\omega^2_\mathrm{BV} = \frac{d\phi}{dr} \left
  (\frac{1}{\gamma}\frac{d\ln{P}}{dr} - \frac{d\ln{\rho}}{dr}\right)\,,
\end{equation}
and take
\begin{equation}
\omega_\mathrm{BV} = \mathrm{sign} (\omega^2_\mathrm{BV}) \sqrt{|\omega^2_\mathrm{BV}|}\,,
\end{equation}
where $d\phi/dr$ is the radial gradient of the gravitational potential
and $\gamma$ is the adiabatic index.  Positive values of
$\omega_\mathrm{BV}$ (blue) denote convectively stable regions while
negative values (red) denote convectively unstable regions. On each
graph we also show the mean and maximum shock radius as green and
magenta lines, respectively, and the average gain radius as a red
line.  While the Brunt-V\"ais\"al\"a frequency shows the susceptibility
to convection, the anisotropic velocity,
\begin{equation}
  v_\mathrm{aniso} = \sqrt{\frac{\langle \rho [(v_r - \langle v_r
      \rangle)^2 + v_\theta^2]\rangle}{\langle \rho \rangle}}\,,
\end{equation}
shows the actual presence of convective motions.  The right panel of
\fref{fig:s20WH07_omega} show the anisotropic velocity in the 2D
simulation of s20.  Also in this panel we show the location of the
heavy-lepton neutrinosphere from the 1D simulation (dashed line) and
from the 2D simulation (solid line).  The neutrinosphere is defined as
the location where the spectrum weighted flux factor equals 0.25.  We
note several important results from these graphs below.

Before discussing PNS convection in detail, we will briefly discuss
the development of convection in the gain region.  The 1D and 2D maps
of $\omega^2_\mathrm{BV}$ are very similar up to $\sim$100\,ms after
bounce. Even though there is a convectively unstable region close to
the shock front, due to the high advection rate of material through
this region convective seeds are unable to grow to full scale
convection before the matter is accreted out of the unstable
region. By $\sim$100\,ms the accretion rate has dropped, the shock
radius increased, and the convectively unstable region has grown so
that convective motions begin to develop in earnest as seen through
the rise of $v_\mathrm{vaniso}$. This causes the shock radius to
deviated from the spherically symmetric solution at this time. The
added pressure support to the material behind the shock from these
anisotropic motions is crucial to aid the neutrino mechanism and help
drive an explosion \citep{couch:15a,mueller:15}.

\fref{fig:s20WH07_omega} also shows the development of a region that
is unstable to PNS convection between $\sim$15\,km-$\sim$25\,km
starting as soon as $\sim$60\,ms after bounce in the 1D model.  While
we just show the GREP model s20, such a region is present in all
models. In the corresponding 2D simulation, the onset of PNS
convection occurs at this time, as seen in the color map of
$v_\mathrm{aniso}$, and works to maintain
$\omega^2_\mathrm{BV} \sim 0$. This convection has a quantitative
effect on the neutrinos being emitted in our 2D simulations. PNS
convection effectively transfers internal energy from deep in the PNS
to slightly larger radii.  This helps support the PNS and prevents the
neutrinospheres from receding as they do in the corresponding 1D
simulations.  This predominantly impacts the heavy-lepton neutrinos as
they are emitted from the deepest regions of the star, closest to the
PNS convection zone.  We explicitly show the heavy lepton
neutrinosphere location in the right panel of
\fref{fig:s20WH07_omega}, the solid line is the neutrinosphere from
the 2D simulations while the dashed line is from the corresponding 1D
simulation.  The increased neutrinosphere radius, and the higher
matter temperatures at the typical radius where the heavy lepton
neutrinos are emitted from gives a larger $\nu_x$ luminosity as can be
seen in \fref{fig:GRnus}. We note that increases, which have also
been attributed to PNS convection, in the $\nu_x$ luminosities are
seen in 1D vs. 2D models as early as \cite{buras:06b} and more
recently in \cite{radice:17}.  However, it is worth mentioning that
\cite{buras:06b} see a relatively smaller impact of PNS convection on
the heavy-lepton neutrino luminosity (an increase of, at most,
$\sim20\%$) than seen here or in \cite{radice:17}.

\section{Conclusion}
\label{sec:conclusion}

The study of core-collapse supernovae is an exceedingly multi-physics
problem.  If anything, decades of research into the core-collapse
supernova central engine has shown that in order to create realistic
simulations of these astrophysical phenomena we need to include all of
the important underlying physics.  This includes, but is of course not
limited to, the nuclear equation of state, neutrino transport and
neutrino interactions, general relativistic gravity, realistic
progenitor models, resolved multidimensional (3D)
magnetohydrodynamics, and more.  In this paper, we have extended the
capabilities of the \code{FLASH} simulation package to now include an
effective treatment of general relativistic gravity and a
multispecies, multigroup, multidimensional, velocity-dependent
neutrino-radiation transport solver.  Both of these new
implementations are designed for application to the core-collapse
supernova problem.

We have tested our GR implementation of gravity and our neutrino
transport solver with several classic test problems, including an
unstable TOV migration test (in both 1D and 2D), and the M1 shadow
test.  Furthermore, we perform the benchmark core-collapse test case
for GR neutrino radiation transport + hydrodynamics from
\cite{liebendoerfer:05} which uses a base set of neutrino interactions
based on \cite{bruenn:85}, the LS180 nuclear EOS from \cite{lseos:91},
and the 15$M_\odot$ progenitor from \cite{ww:95}.  Our results closely
match those of other neutrino transport methods. We critically assess
and make clear the impact of the approximations in our neutrino
transport scheme with this comparison.

Extending our simulations to axisymmetry, we study core collapse in
four progenitors from \cite{woosley:07} with both Newtonian and GR
gravity for at least $\sim$800\,ms after bounce.  While the benefits
that a general relativistic treatment of gravity can have for the
core-collapse supernova problem have been discussed before, ours is
the first study to systematically explore the differences between
Newtonian and GR gravity in multiple dimensions and across several
updated progenitor models.  Our results clearly and unambiguously show
that GR gravity significantly aids the neutrino mechanism in
reenergizing the stalled supernova shock formed in the core collapse
of massive stars. We obtain explosions in three of four progenitor
models from \cite{woosley:07} when using general relativistic
gravity. Two of the four corresponding Newtonian simulations also
achieve explosions, however they are delayed by an additional $\sim$500\,ms.
There are numerous examples in the literature of complicated multi-physics
simulations of core-collapse supernova with sophisticated neutrino
treatments but only Newtonian gravity. The results presented here show that
the approximation of Newtonian gravity can be quite detrimental in the
context of core-collapse supernova explosions and should be relaxed if
accurate results are sought.

We have also showed the stochastic nature of the core collapse problem
in 2D axisymmetry by simulating one model with 5 different sets of
random initial perturbations.  For this model, we find the explosion
time can vary by $\pm \sim$100\,ms. Such stochastic variations must be
considered when making detailed comparisons between simulations.  We
have also added four additional models to our GREP simulations, s21,
s22, s23, and s24 from \cite{woosley:07}.  This region is predicted
from several 1D parameterized explosions studies to be susceptible to
black hole formation.  We do not see any evidence of this, instead we
find explosions in three of four of these models.

Our results match, in several ways, other multidimensional,
energy-dependent neutrino radiation transport simulations of
core-collapse supernovae including the Newtonian gravity simulations
of \cite{dolence:15} and the GR gravity simulations of
\cite{summa:16}.  However, differences still exist between
the results of these works and the work presented here.  This could be
due to differences in the neutrino and nuclear microphysics, neutrino
transport methods, and/or numerical setups.  We find qualitative
differences between our work and the work of
\cite{bruenn:13,bruenn:16}, which again, could be due to any of
the potential differences listed above.  We note specifically 
significant differences in the amount of the neutrino heating present
during the early postbounce phase.  A close comparison of methods,
input physics, and results is warranted to resolve these differences.

\section*{Acknowledgements}

The authors acknowledge fruitful conversations with F. Foucart,
J.~A. Harris, R. Hix,
H.~T. Janka, E. Lentz, O.~E.~B. Messer, C.~D. Ott, L.~F. Roberts. Computations
were performed on the Zwicky cluster at Caltech, which is supported by
the Sherman Fairchild Foundation and by NSF award PHY-0960291.
Computations were also performed on XSEDE which is
supported by National Science Foundation grant number ACI-154856 and the gpc supercomputer at the
SciNet HPC Consortium. SciNet is funded by: the Canada Foundation for
Innovation under the auspices of Compute Canada; the Government of
Ontario; Ontario Research Fund - Research Excellence; and the
University of Toronto \citep{loken:10}.
An award of computer time was provided by the Innovative and Novel Computational Impact on Theory and Experiment (INCITE) program. This research used resources of the Argonne Leadership Computing Facility, which is a DOE Office of Science User Facility supported under Contract DE-AC02-06CH11357.
Support for this work was
provided by NASA through Hubble Fellowship grant \#51344.001-A awarded
by the Space Telescope Science Institute, which is operated by the
Association of Universities for Research in Astronomy, Inc., for NASA,
under contract NAS 5-26555.

\software{FLASH \citep{fryxell:00, dubey:09}, GR1D \citep{oconnor:10,oconnor:15a}, 
                              Matplotlib
                              \citep{matplotlib}, VisIt
                              \citep{HPV:VisIt} }

\appendix

\section{Neutron Star Migration Test}
\label{app:GR}

\begin{figure}[ht]
\centering
\includegraphics[width=0.5\columnwidth]{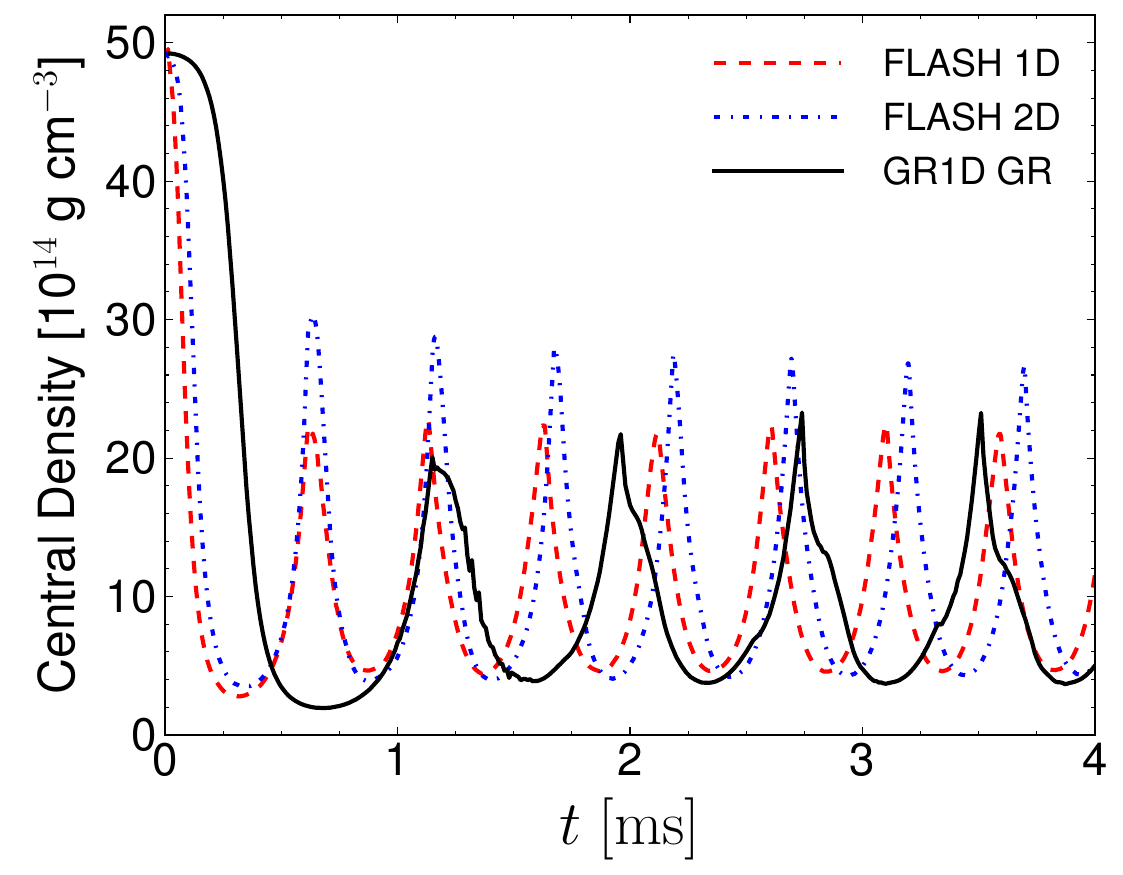}
\caption{Unstable TOV star migration test.  This TOV star migration
  test is performed in \code{FLASH} to test our implementation of the
  `case A'  GR effective potential from \cite{marek:06}.  We present
  results from a 1D simulation (red dashed line) a 2D simulation (blue
dashed-dotted line) and a full GR simulation in \code{GR1D} (black
solid line). The results compare quite well to those of
\cite{marek:06} and convince us of the correctness of our effective
potential implementation. }\label{fig:migration}
\end{figure}

In \fref{fig:migration}, we test our GR effective potential
implementation via a Tolman-Oppenheimer-Volkoff (TOV) star migration
test. We follow the initial conditions from \cite{marek:06}.  Our
unstable TOV stars follow a $\Gamma=2$ polytropic EOS,
$P=K\rho^\Gamma$ with a value of $K=1.455\times 10^5$ [cgs]. We take
an initial central density of $\rho_c=4.93\times10^{15}$\,g\,cm$^{-3}$
and solve the `case A' modified TOV equations to obtain the
equilibrium (but unstable) structure.  With simulate 4\,ms of
evolution in \code{FLASH}, in both 1D (red dashed line) and 2D (blue
dashed-dotted). We also use the full GR code \code{GR1D} to perform
the migration test (black solid line). For this full GR test, we use
the standard TOV equations to construct the initial profile. In the
effective potential tests, the unstable star quickly ($\sim$0.1\,ms)
starts its migration away from the unstable configuration, and
oscillates about the stable configuration. The full GR evolution is
slower, especially at the beginning, which we can understand from the
time dilation effect in the full GR simulation, the value of the
central lapse for this configuration begins at $\sim$0.27, and
migrates to $\sim$0.65. Our results compare very well to the `case A'
results and the full GR \code{CoCoNut} results of \cite{marek:06},
completing between 7 and 8 oscillation cycles in the first 4\,ms of
evolution for the effective potentials and $\sim$5 cycles in the full
GR case.  This convinces us that we have correctly implemented the GR
effective potential in both 1D and 2D in \code{FLASH}.  We have also
tested stable TOV star configurations. Using a stable TOV star, with the
same EOS and an initial central density of
$7.5\times10^{14}$\,g\,cm$^{-3}$, \code{FLASH} maintains this central
density over 4\,ms to better than 1\% in 1D and 0.5\% in 2D.

\section{Neutrino Transport Methods \& Implementation}
\label{app:code}

In this appendix we present the details of the numerical techniques
used to solve the neutrino transport equations (\eref{eq:Esph} to
\eref{eq:Fzcyl}), the microphysics used to compute the neutrino
interaction coefficients, and the couple to the hydrodynamics.

\emph{Closure:} To close the hierarchy of moment evolution equations
after the first two moments, we must specify a closure relation for
the Eddington tensor $P^{ij}$ in terms of the two lower moments $E$
and $F^{i}$. We choose the common M1 closure and perform the
calculation in the fluid rest frame in order to achieve the
appropriate limiting cases. Our framework follows closely that of
\cite{oconnor:15a,shibata:11}. In the optically thick regime
the radiation in the fluid rest frame is isotropic.  For this limit, we
write the Eddington tensor as
$K^{ij} \equiv K_\mathrm{thick}^{ij} = J/3 h^{ij}$, where
$h^{ij}=1+W^2v^iv^j$ is the projection tensor. In optically thin, free
streaming regions far from the source
$K^{ij} \equiv K_\mathrm{thin}^{ij} = J (H^iH^j/H^2)$.  For the total
Eddington tensor in the fluid rest frame we interpolate between these
two limiting regimes,
\begin{equation}
K^{ij} = \frac{3(1-\chi)}{2}K_\mathrm{thick}^{ij} + \frac{3\chi-1}{2}K_\mathrm{thin}^{ij}\,.\label{eq:closure}
\end{equation}
In these equations $\chi$ is taken \citep{minerbo:78} to be
\begin{equation}
\chi = \frac{1}{3} + \frac{2}{15}(3f^2 - f^3 + 3f^4)\,,
\end{equation}
where $f \equiv (H^\alpha H_\alpha / J^2)^{1/2}$ is the flux
factor. $f$ is equal to 0 if the radiation is isotropic, which gives a
$\chi=1/3$ and $K^{ij} = K_\mathrm{thick}^{ij}$. $f$ is 1 if the
radiation is fully forward peaked in some direction. For this case,
$\chi=1$ and $K^{ij} = K_\mathrm{thin}^{ij}$. To get the Eddington
tensor in the laboratory frame, $P^{ij}$, we take the spatial
components of the neutrino stress energy tensor constructed from the
fluid frame values,
$P^{ij} = T^{ij} = JW^2v^iv^j + H^iWv^j + H^jWv^i + K^{ij}$. In
practice, we compute $J$ and $H^\alpha$ via
\eref{eq:fluidframeJ}-\eref{eq:fluidframehi} using a guess for
$P^{ij}$ constructed assuming zero velocity, we then determine
\eref{eq:closure}, and finally $P^{ij}$.

\emph{Explicit Fluxes:} The flux terms, both spatial and energy-space,
on the left hand side of \eref{eq:Esph}-\eref{eq:Fzcyl} are computed
using the value of the radiation field at the beginning of the time
step. For the energy-space fluxes we follow the methods of
\cite{mueller:10}. For the energy-space flux term, we use the
slow-motion limit expressions from \cite{shibata:11} and the
gravitational redshift terms from \cite{rampp:02}.  Specifically, we
take,
\begin{equation}
L^{ij} = \frac{3\chi-1}{2} \frac{E\,
  F^i\,F^j}{F^kF_k}+\frac{1-\chi}{2}E\delta^{ij}\,,
\end{equation}
and
\begin{equation}
N^{ijk}\partial_jv_k =  \frac{3\chi-1}{2} \frac{E\, F^i\,F^j\,
  F^k}{(F^lF_l)^{3/2}}\partial_jv_k + \frac{3(1-\chi)}{10}\left(F^i\partial_jv^j+F^j\partial_jv^i+F^j\partial_iv_j \right)\,.
\end{equation}

The spatial fluxes are computed using an approximate Riemann solver
with corrections for the optically thick regime. We describe this in
detail below. To begin, we write the spatial fluxes as a finite
difference between the flux at each interface,
\begin{equation}
\partial_x [ \alpha r^m F^x ] = \frac{(\alpha r^m)_{k+} \mathcal{F}_{k+} -
  (\alpha r^m)_{k-} \mathcal{F}_{k-}}{\Delta x}\,,
\end{equation}
and
\begin{equation}
\partial_x [\alpha r^m P^{xi} ] =\frac{ (\alpha r^m)_{k+}
  \mathcal{P}^i_{k+} - (\alpha r^m)_{k-} \mathcal{P}^i_{k-}}{\Delta x}\,,
\end{equation}
where $m$ is either $0,1,2$ and we use $k\pm$ to denote the $k\pm1/2$
interface.  To obtain the interface fluxes $\mathcal{F}_{k+}$ and
$\mathcal{P}^i_{k+}$, we use the standard HLLE Riemann solver for
hyperbolic equations. For the flux evaluation at an interface $k\pm$,
we reconstruct $E$ and $F^i/E$ to both sides of the zone interfaces
using 2nd-order TVD (total variation diminishing) interpolation.  On
both sides of the interface we recompute the closure as described
above to obtain the cell interface values of $P_{ij}$.  The
characteristic speeds for the Riemann solver are calculated in a
similar way as the closure in that we interpolate between the
optically thick and optically thin limits. First, for each interface
(characterized here by the direction $k$), we determine the minimum
and maximum speeds on each side of the interface in both the optically
thin and optically thick limits. For the optically thick limit the
choice is clear,
\begin{equation}
\lambda^{(k)}_\mathrm{thick,min} = -\frac{1}{\sqrt{3}}; \lambda^{(k)}_\mathrm{thick,max} = +\frac{1}{\sqrt{3}}\,.
\end{equation}
For the thin limit, the maximum and minimum characteristic speeds are
\citep{shibata:11}
\begin{equation}
\lambda^{(k)}_\mathrm{thin,min/max} = \mathrm{min/max}\left(\frac{\pm F^{k}}{\sqrt{F_i F^i}},E\frac{F^k}{F_iF^i}\right).
\end{equation}
Next, to determine the maximum and minimum speed on each side of the
interface we interpolate between the optically thick
($\lambda_\mathrm{thick}^{(k)}$) and free streaming
($\lambda^{(k)}_\mathrm{thin}$) regimes via,
\begin{equation}
\lambda^{(k)}_\mathrm{min/max} = \frac{3(1-\chi)}{2}\lambda^{(k)}_\mathrm{thick, min/max}+ \frac{3\chi-1}{2}\lambda^{(k)}_\mathrm{thin, min/max}\,.
\end{equation}
The final step to determine the minimum and maximum speeds for the
Riemann solver is to take
$\lambda^{(k)+} =
\mathrm{max}(\lambda^{(k),(R)}_\mathrm{max},\lambda^{(k),(L)}_\mathrm{max})$
and
$\lambda^{(k)-} =
\mathrm{min}(\lambda^{(k),(R)}_\mathrm{min},\lambda^{(k),(L)}_\mathrm{min})$.
Where $(R)$ and $(L)$ denote the right and left side of the interface,
respectively.

With the reconstructed moments and minimum and maximum characteristic
speeds in hand, the HLLE Riemann solution for the fluxes at the
interface is then,
\begin{equation}
\mathcal{F}_{k+,\mathrm{HLLE}} = \frac{\lambda^{(k)+} F^{k,(L)} - \lambda^{(k)-} F^{k,(R)} + \lambda^{(k)+} \lambda^{(k)-} (E^{(R)}-E^{(L)})}{\lambda^{(k)+}-\lambda^{(k)-}}
\end{equation}
and
\begin{equation}
\mathcal{P}^{j}_{k+,\mathrm{HLLE}} = \frac{\lambda^{(k)+} P_{kj}^{(L)} - \lambda^{(k)-} P_{kj}^{(R)} + \lambda^{(k)+} \lambda^{(k)-} (F_j^{(R)} - F_j^{(L)})}{\lambda^{(k)+}-\lambda^{(k)-}}
\end{equation}
where here $(R)$ and $(L)$ label the reconstructed (or recalculated for
$P^{ij}$) moments on the right and left side of the interface $k+1/2$,
respectively.

However, this solution relies on the equations being hyperbolic. This
condition is violated in regions with high opacity and numerical
diffusion can dominant the HLLE Riemann solution.  The transition away
from hyperbolicity can be traced via the Peclet number
($\mathrm{Pe} \equiv \Delta x (\kappa_s+\kappa_a)$).  If Pe is
$\gtrsim1$ then we must correct the Riemann solution.  We follow
\cite{audit:02,oconnor:13} and ultimately take the following for the
interface fluxes,
\begin{equation}
\mathcal{F}_{k+} = a \times
\mathcal{F}_{k+,\mathrm{HLLE}} + (1-a) \times \mathcal{F}^{(k)}_\mathrm{asymptotic}\,,
\end{equation}
where $a \equiv \mathrm{tanh}(1/\overline{\mathrm{Pe}})$ with
$\overline{\mathrm{Pe}}$ is the geometric mean of the Peclet numbers on
either side of the interface and,
\begin{equation}
\mathcal{F}^{(k)}_\mathrm{asymptotic} = \frac{4}{3} (v^k
J)_\mathrm{upstream} - \frac{1}{3\bar{\kappa}}\frac{J^{(k+1)}-J^{(k)}}{\Delta x_k}\,.
\end{equation}
For the momentum density spatial flux we use,
\begin{equation}
\mathcal{P}^{j}_{(k+1/2)} = a \times
\mathcal{P}^{j}_{k+,\mathrm{HLLE}} + (1-a) \times \mathcal{P}^{j}_\mathrm{k+,asymptotic}\,,
\end{equation}
where
\begin{equation}
\mathcal{P}^{j}_\mathrm{k+,asymptotic} = (P_{kj}^{(R)} + P_{kj}^{(L)})/2
\end{equation}

In Appendix \ref{app:shadow}, we perform a 2D shadow test to ensure the correct
multidimensional implementation of our transport scheme.  There we
show one of the beneficial features of M1, the ability to maintain
neutrino momentum in the free streaming regime.  In one moment
transport schemes, like flux limited diffusion, the flux of neutrinos
is determined by the spatial gradient of the zeroth moment.  In the
context of core-collapse supernovae, this can cause a smoothing out of
the radiation field at large distances \citep{ott:08}.

\emph{Neutrino Interactions:} The last aspect of the neutrino moment
evolution equations (\eref{eq:Esph}-\eref{eq:Fzcyl}) are the neutrino
interaction coefficients.  These coefficients are interpolated from a
table computed via {\tt{NuLib}} \citep{oconnor:15a}, an open-source
neutrino interaction library. We include a basic set of neutrino
interactions and study three independent neutrino species: electron
neutrino $\nu_e$, electron antineutrino $\bar{\nu}_e$, and a
characteristic heavy-lepton neutrino
$\nu_x = \{\nu_\mu,\bar{\nu}_\mu,\nu_\tau,\bar{\nu}_\tau\}$. For
electron neutrino and electron antineutrino emission processes, we
include electron and positron capture on protons and neutrons,
respectively following \cite{bruenn:85,brt:06} and include weak
magnetism and nucleon recoil corrections from \cite{horowitz:02}. We
also include electron neutrino emission via electron capture on heavy
nuclei via the simple formalism of \cite{bruenn:85}. Neutrino
emissivities are computed from the absorption opacities via
Kirchhoff's law which equates the emissivity to the rate of absorption
of an equilibrium neutrino distribution:
$\eta = \kappa_a E_\mathrm{equil}$.  Heavy-lepton neutrino emission
from electron-positron annihilation and nucleon-nucleon Bremsstrahlung
is handled via an approximation which computes an effective emissivity
and absorption opacity and does not require energy group coupling.
This approach works well for the core collapse problem
\citep{oconnor:15a}.  We include elastic scattering on neutrons and
protons, which also includes weak magnetism and recoil corrections,
elastic scattering on alpha particles and on heavy nuclei, with the
latter including corrections for the heavy-ion form factor, ion-ion
correlations, and electron polarization \citep{brt:06}.

\emph{Hydrodynamic coupling:} During each time step we solve the
neutrino moment evolution equations operator-split from the
hydrodynamics. We first update the hydrodynamic variables from time
step $(n)$ to $(n+1)$ using {\tt{FLASH}}'s unsplit hydrodynamics
solver, fifth-order reconstruction with WENO, and the HLLC Riemann
solver. We then use the updated matter variables
($\rho^{(n+1)}$,$T^{(n+1)}$, and $Y_e^{(n+1)}$) to compute the $(n+1)$
neutrino-matter interaction coefficients, $\eta^{(n+1)}$,
$\kappa_a^{(n+1)}$, and $\kappa_s^{(n+1)}$. Our neutrino radiation
evolution methods (first order time stepping and second order spatial
reconstruction) in concert with the four guard cells needed for the
hydrodynamic evolution affords us the possibility to evolve two
radiation substeps for each hydrodynamic step. This greatly reduces
the communication needed which will be a limiting issue in simulations
involving three dimensions. For the explicit part of each of the two
radiation substeps we use the most recent radiation field variables
(either the $(n)$ or $(n+1/2)$ values) to determine the fluxes (both
spatial and energy-space) as described above. For each radiation
substep, these fluxes are treated as an explicit source term.  The
remaining terms, which are all local to each grid zone and only couple
the neutrino energy density and momentum density of a given group and
species, are solved via implicit integration to arrive at
$E^{(n+1/2)}$ and $F^{i,(n+1/2)}$ after the first substep and 
$E^{(n+1)}$ and $F^{i,(n+1)}$ after the second. For these substeps we 
assume $P^{ij,(n+1/2)} = p^{ij,(n)} E^{(n+1/2)}$ and 
$P^{ij,(n+1)} = p^{ij,(n+1/2)} E^{(n+1)}$. We compute the energy, 
momentum, and lepton exchange with the matter for each substep and 
update $v_i^{(n+1)}$, $e^{(n+1)}$ and $Y_e^{(n+1)}$ at the end of the 
radiation step. For each substep, the change in $\rho v_i$ is taken as 
\begin{equation}
\Delta [\rho v_i] = -4\pi \alpha \Delta t \sum_{\nu, \epsilon_j}
   \Delta \epsilon \left [ -(\kappa_s + \kappa_a)H^i +
  W(\eta - \kappa_a J)v^i\right]\,, 
\end{equation}
where the sum is over all neutrino species and neutrino energy groups. 
The change in the total specific energy, $e_\mathrm{tot}$, is taken as 
\begin{equation}
\Delta [\rho e_\mathrm{tot}] = \Delta [\rho e_\mathrm{int}] + \Delta 
   [\rho e_\mathrm{kin}]
= -4\pi \alpha \Delta t \sum_{\nu, \epsilon_j}
   \Delta \epsilon \left [
  W(\eta - \kappa_a J) - (\kappa_a + \kappa_s)H^t\right]\,. 
\end{equation}
In practice, we update $v_i$, $e_\mathrm{tot}$, then 
determine $e_\mathrm{int}$ for use in the EOS update by subtracting 
off the new $e_\mathrm{kin}$.  We explicitly mention this because in 
a previous iteration of this paper \citep{oconnor:15b_v1} we had 
mistakenly ignored the contribution to the total energy due to the 
changing momentum. The change in $Y_e$ is taken as 
\begin{equation}
\Delta [\rho Y_e] =  -4\pi \alpha m_\mathrm{amu} \Delta t \sum_{\nu, 
  \epsilon_i} \Delta \epsilon \left [
  s_\nu \frac{(\eta - \kappa_a J)}{\epsilon} \right]\,, 
\end{equation}
where $s_\nu$ is +1 for electron neutrinos, -1 for electron 
antineutrinos, and 0 for heavy-lepton neutrinos.  The use of an 
operator split method over a coupled method between the radiation and 
the hydrodynamics is justified because of the small time step imposed 
by the explicit update of the neutrino radiation fields.

\section{M1 Shadow Test}
\label{app:shadow}

In one-moment transport schemes, like flux-limited diffusion, the flux
of energy density is taken to be in the direction of the spatial
gradient of the neutrino energy density.  This works well in the
diffusion limit, but in a free streaming regime this assumption works
to artificially source neutrino momentum and wash out potentially
physical gradients.  One of the advantages of a M1-like transport
scheme is the ability to maintain the momentum of free streaming
neutrinos.  The classic test problem to demonstrate this ability of M1
transport is the shadow test. We closely follow the 2D shadow test
proposed in \cite{just:15b}, with the exception that we employ
axisymmetry and use cylindrical coordinates whereas \cite{just:15b}
perform their shadow test in genuine 2D Cartesian coordinates, apart
from a geometric factor of $r$, these setups are identical.  At the
origin, within a spherical radius of $r_s= 1.5$, we have a spherical
emission source with an absorption opacity of
\begin{equation}
\kappa_a = 10 \exp{[-(4r/r_s)^2]}\,,
\end{equation}
and an emissivity with an arbitrary normalization.  Along the radial
direction, a distance of 8 from the central source, we place a purely
absorbing ($\kappa_a = 10$) circle (which corresponds to a torus under
axisymmetry) with radius of 2. The grid extends to +12 in the radial
direction and $\pm$5 along the symmetry axis.  The grid resolution is
0.05. For this test we do not use any mesh refinement, a side effect
of this is that we have free streaming radiation on the finest
resolution level of the grid and the first order method for the explicit flux
calculation does poorly unless we lower the CFL factor from 0.4 to
0.2, which we do for this test.  We do not encounter this issue with
our core collapse simulations because of our mesh refinement, the
neutrino radiation is not close to the free streaming limit until $r
\gtrsim 100$\,km, by that radius we have decreased our grid
resolution, giving small effective CFL factors in these outer zones.

\begin{figure}[ht]
\centering
\includegraphics[trim= 0cm 0cm 0cm 0cm, clip,  width=0.5\columnwidth, angle=270]{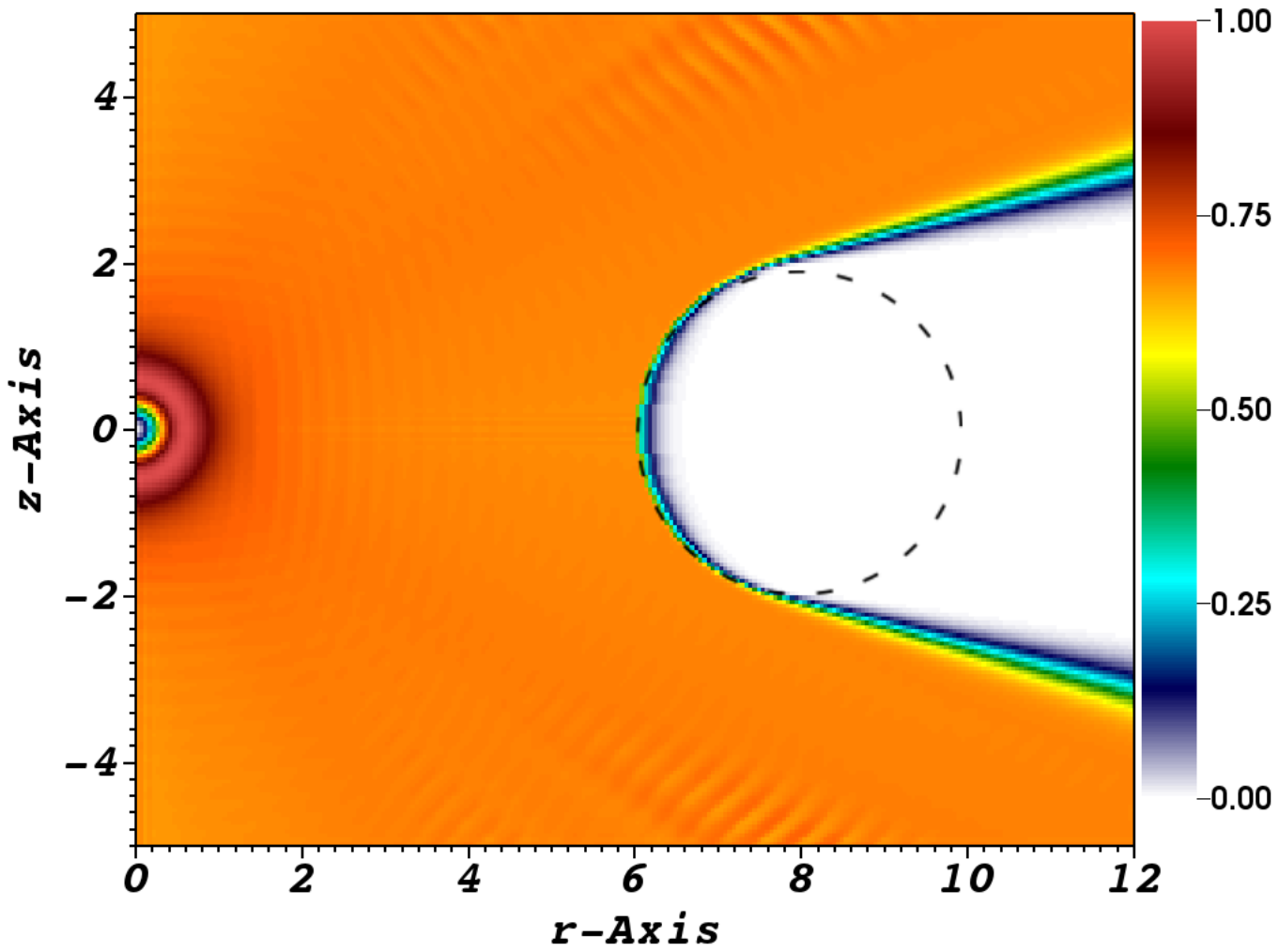}
\caption{Neutrino energy density multiplied by $r^2$ in our M1 shadow
  test in 2D cylindrical coordinates. There is a spherical emission
  source located at the origin and a perfectly absorbing region
  (marked by the dashed circle) at $r=8$ with a radius of 2. This test
  closely follows the setup of \cite{just:15b}.}\label{fig:shadow}
\end{figure}

In \fref{fig:shadow}, we show the radiation transport solution for the
neutrino energy density.  To achieve a constant value in the free
streaming regime we scale the energy density by $r^2$.  We normalize
the data to range between 0 and 1.  We show the solution after several
light crossing times of the domain.  The shadow cast by the absorbing
torus is clearly seen and demonstrates this ability of our code to
maintain the neutrino momentum direction. There are slight artifacts
seen as ripples at larger radii and along the coordinate axes nearer
to the source that arising from the first-order explicit spatial flux
integration.

\end{document}